\documentclass[11pt]{article}
\pdfoutput=1

\usepackage{cite}
\usepackage{amsmath,amssymb,amsbsy,amstext, amsthm, simplewick}
\usepackage{hyperref}
\usepackage{graphicx}
\usepackage{amsfonts}
\usepackage{amssymb}
\usepackage[small]{caption}
\usepackage{upgreek}
\usepackage[svgnames,dvipsnames,x11names,table]{xcolor}
\usepackage{multirow} 
\usepackage{geometry}
\usepackage[hang,flushmargin]{footmisc}
\usepackage{bm}
\usepackage{braket}
\usepackage{subcaption}
\usepackage{simpler-wick}
\usepackage{mathtools}
\usepackage{slashed}

\usepackage{colortbl}

\makeatletter
\newlength{\apb@width}
\newcommand{\autoparbox}[2][c]{\settowidth{\apb@width}{#2}\parbox[#1]{\apb@width}{#2}}

\makeatother

\usepackage[most]{tcolorbox}
\tcbset{colback=white, colframe=black,
        highlight math style= {enhanced, 
            colframe=red,colback=red!10!white,boxsep=0pt}
        }
\definecolor{light-gray}{gray}{0.95}

\definecolor{lightgray}{gray}{0.9}

\usepackage[framemethod=default]{mdframed}
\newmdenv[skipabove=7pt,
skipbelow=7pt,
rightline=false,
leftline=false,
topline=false,
bottomline=false,
backgroundcolor=gray!10,
linecolor=gray,
innerleftmargin=5pt,
innerrightmargin=5pt,
innertopmargin=5pt,
innerbottommargin=5pt,
leftmargin=0cm,
rightmargin=0cm,
linewidth=4pt]{eBox}

\numberwithin{equation}{section}

\def\beq{\begin{equation}}
\def\eeq{\end{equation}}

\def\bea{\begin{eqnarray}}
\def\eea{\end{eqnarray}}

\def\beq{\begin{equation}}
\def\eeq{\end{equation}}
\def\bea{\begin{eqnarray}}
\def\eea{\end{eqnarray}}

\def\Mpl{M_{\rm pl}}
\def\Mp{M_{\rm pl}}

\def\k{{\vec k}}

\DeclareRobustCommand{\SkipTocEntry}[4]{}

\definecolor{blue3}{RGB}{31, 119, 180}
\definecolor{red3}{RGB}{	214, 39, 40}
\definecolor{orange3}{RGB}{255, 127, 14}
\definecolor{green3}{RGB}{44, 160, 44}

\newcommand{\Fig}[1]{Fig.~\ref{#1}}
\newcommand{\Eq}[1]{Eq.~\eqref{#1}}
\newcommand{\Eqs}[2]{Eqs.~\eqref{#1} and \eqref{#2}}

\newcommand{\nn}{\nonumber}
\newcommand{\delC}{\delta_c}

\newcommand{\order}{\mathcal{O}}

\newcommand{\JBG}{\mathcal{J}}
\newcommand{\Lag}{\mathcal{L}}
\newcommand{\amp}{\mathcal{A}} 
\newcommand{\prop}{\mathcal{D}}
\newcommand{\Kmom}{\hat{\bm K}}
\newcommand{\cm}{c_{\phi}}
\newcommand{\delPhi}{\delta_\phi}
\newcommand{\pp}{\tilde{t}}

\begin{document}

\begin{titlepage}
\setcounter{page}{1} \baselineskip=15.5pt 
\thispagestyle{empty}

\begin{center}
{\fontsize{18}{18} \bf Inflationary Adler Conditions }
\end{center}

\vskip 20pt
\begin{center}
\noindent
{\fontsize{12}{18}\selectfont  Daniel Green, Yiwen Huang, and Chia-Hsien Shen}
\end{center}

\begin{center}
\textit{ Department of Physics, University of California at San Diego, \\ La Jolla, CA 92093, USA}
\end{center}

\vspace{0.4cm}
 \begin{center}{\bf Abstract}
  \end{center}

We derive a new soft theorem that corresponds to the spontaneous breaking of Lorentz boosts.
This is motivated by the dynamics of inflation in the sub-horizon (flat-space) limit, where 
spacetime becomes flat but Lorentz boosts are still broken.
In this limit, the scattering amplitudes become sensible observables.
We relate the soft emission of a Goldstone boson to the (non-relativistic) Lorentz boost of the hard scattering amplitudes.
This is the on-shell avatar of the spontaneous breaking of Lorentz boosts,
analogous to the Adler zero of pions in the chiral symmetry breaking.
We comment on several applications to inflation, including the demonstration that Dirac-Born-Infeld Inflation is the unique theory that has an emergent Lorentz invariance when the boosts are spontaneously broken.

\end{titlepage}

\restoregeometry

\newpage
\setcounter{tocdepth}{2}
\tableofcontents

\newpage

\section{Introduction}

Inflation is widely believed to be an essential part of the history of the universe.  It explains numerous features of the universe we observe; significantly, the initial seeds of structure are believed to have formed during this era from quantum vacuum fluctuations. Observational tests of the inflationary epoch are limited to the statistics of fluctuations produced during inflation, including scalar and tensor metric fluctuations.  However, learning concrete lessons from these fluctuations is limited by our ability to relate the space of models of inflation with the space of consistent statistical correlations.

Effective field theory (EFT) provides a natural framework in which we can relate microscopic theories with long distance observables. The EFT of Inflation\cite{Creminelli:2006xe,Cheung:2007st} is such an example where one can write a theory for the fluctuations directly, where the microscopic details are encoded in the particle content and couplings.  For single-field inflation specifically, the lone scalar degree of freedom is the scalar metric fluctuation and is highly constrained by symmetries~\cite{Maldacena:2002vr,Creminelli:2004yq}. For scale-invariant fluctuations, the dynamics of the inflationary era are encoded in higher-dimensional operators whose coefficients are constrained by primordial non-Gaussianity in the cosmic microwave background~\cite{Planck:2019kim} or distribution of galaxies~\cite{Cabass:2022wjy,DAmico:2022gki} (see~\cite{Achucarro:2022qrl} for a recent review).

In many contexts, it has been found that the space of EFTs consistent with our short distance symmetries is significantly smaller than our naive EFT expectations suggest~\cite{deRham:2022hpx}.  This has been seen in countless examples through the self-consistency of the scattering amplitudes calculated within the EFT~\cite{Pham:1985cr,Adams:2006sv}.  Naively, such an approach does not apply to the EFT of Inflation which is defined in quasi-de Sitter space.  However, ongoing work on the cosmological bootstrap~\cite{Baumann:2022jpr} has shown how cosmological correlators in inflation are tied to the scattering amplitudes of the same theory in flat space~\cite{Maldacena:2011nz,Raju:2012zr,Arkani-Hamed:2015bza,Lee:2016vti,Arkani-Hamed:2018kmz,Arkani-Hamed:2018bjr,Benincasa:2018ssx,Baumann:2019oyu,Pajer:2020wxk,Baumann:2020dch,Bonifacio:2021azc,Cabass:2021fnw,Baumann:2021fxj,Benincasa:2022gtd}.  In this precise sense, understanding how amplitudes arise within the flat-space limit of the EFT of Inflation is directly related to our understanding of cosmic observables. Preliminary work in this direction has been initiated in~\cite{Baumann:2011su,Baumann:2015nta,Baumann:2019ghk,Pajer:2020wnj,Grall:2021xxm,Creminelli:2022onn}, but many questions about the general structure of the amplitudes remain.  
The most natural context to discuss the flat-space limit of the EFT of Inflation is the so-called decoupling limit~\cite{Baumann:2011su}. In this regime, dynamical gravity is decoupled from inflation and the scalar metric mode is described by the Goldstone boson, $\pi$, associated with spontaneously broken time diffeomorphism. The interactions of $\pi$ individually break Lorentz boosts but non-linearly realize the symmetry. Much of this structure also survives in cosmological correlators as the scalar metric fluctuation, $\zeta$, at horizon crossing are well described by the Goldstone boson $\zeta \approx - H \pi$.

To flesh out the the full constraints on inflation from scattering amplitudes in the decoupling limit, it is imperative to understand the full on-shell consequences of spontaneous breaking of Lorentz boosts. The nature of the non-linearly realized symmetry is often tied to the soft limit of the amplitude. A celebrated example is the Adler zero of the soft pion~\cite{Adler:1964um} that reflects the underlying chiral symmetry breaking~\cite{Cronin:1967jq,Weinberg:1966fm,Weinberg:1968de,Coleman:1969sm,Callan:1969sn}.
This approach has been revived recently to identify EFTs from an on-shell perspective~\cite{Cheung:2014dqa,Huang:2015sla,DiVecchia:2015jaq,Padilla:2016mno,Low:2017mlh,Cheung:2018oki,Low:2018acv,Kampf:2019mcd,Kampf:2020tne,Kampf:2021bet,Kampf:2021tbk}. In fact, these soft limits provide powerful means to reconstruct  amplitudes~\cite{Cheung:2015ota,Luo:2015tat,Arkani-Hamed:2016rak,Rodina:2016jyz,Rodina:2018pcb,Bartsch:2022pyi} and classify the space of EFTs~\cite{Cheung:2016drk,Cheung:2018oki,Elvang:2018dco}.
On-shell methods have been applied to non-relativistic EFTs~\cite{Brauner:2020ezm,Pajer:2020wnj,Stefanyszyn:2020kay,Grall:2020ibl,Mojahed:2021sxy,Mojahed:2022nrn,Brauner:2022ymm} and wave functions~\cite{Bittermann:2022nfh}.
For the non-linearly realized Lorentz boosts, which are of central interest to inflation, the  Goldstone theorem has been derived~\cite{Alberte:2020eil} and the applications to conformal field theory has been studied~\cite{Komargodski:2021zzy}, but the analogous soft theorem is yet to be formulated.

In this paper, we will establish a soft theorem for flat-space scattering amplitudes that reflects the spontaneous breaking of Lorentz boosts. 
Our soft theorem, summarized in Section~\ref{sec:softthm_summary}, is the flat-space analog of the sub-leading consistency condition on correlators~\cite{Creminelli:2012ed}.
The on-shell soft theorem also complements the algebraic constructions of boost-breaking EFTs~\cite{Grall:2020ibl}.
Our results will be helpful in understanding cosmological correlators in several ways. 
First, it has been observed that Dirac-Born-Infeld (DBI) Inflation has an emergent Lorentz invariance in the broken phase~\cite{Grall:2020ibl}.
We will use the soft theorem to show that DBI Inflation is the unique model (to leading derivative expansion) that has this property.
Second, it was found in~\cite{Pajer:2020wnj} that boost-violating amplitudes are severely constrained by consistency of the scattering amplitudes (when coupled to gravity). These results seem inconsistent with the EFT of Inflation and the authors suggested it was a result of an inconsistency of the EFT of Inflation in flat space away from the decoupling limit.  Our investigation will show the origin of these constraints and how they do not arise when the non-linear structure of the EFT is enforced.  Finally, we will see how the structure of on-shell observables in flat space has nontrivial implications for inflationary correlators.

The paper is organized as the following. In Section~\ref{sec:review}, we review the EFT of Inflation under the decoupling limit, with emphasis on the non-linearly realized Lorentz boosts and dependence on the field basis. The main results of this paper are given in Section~\ref{sec:softthm}. After reviewing the necessary tools, we prove the soft theorem for Goldstone-boson scattering, the full general case with matter interaction, and comment on the non-perturbative validity of the soft theorem. For readers' convenience, we summarize the final soft theorem in Section~\ref{sec:softthm_summary}.
In Section~\ref{sec:inflation}, we discuss possible applications to inflation. Final conclusion and future directions are discussed in Section~\ref{sec:conclusions}.

\paragraph{Convention:}
We use the metric with mostly minus signature throughout the paper and set the speed of light $c=1$.
The Greek and Roman indices denote components of relativistic and spatial vectors. 
We use boldface $\bm v$ for a spatial vector. 
For the scattering amplitudes, all momenta are outgoing. 
We use the hard momentum $p_a^\mu = (E_a, \bm p_a)$ for particle $a$ and the soft momentum $q^\mu = (\omega, \bm q)$.
We define a rescaled inner product for general $c_s$ using \Eq{eq:p_def}.
The deviation of the propagation speed from the speed of light are defined as $\delC = c_s^{-2}-1$ and $\delPhi = c_\phi^{-2}-1$ for Goldstone boson $\pi$ and a matter field $\phi$, respectively. 

\section{The EFT of Inflation in Flat Space}
\label{sec:review}

\subsection{Action}
The EFT of single-field inflation describes the fluctuation of the inflaton field around a flat Friedmann–Lemaitre–Robertson–Walker (FLRW) background. Time diffeomorphism, $t\rightarrow t+\xi$, is spontaneously broken by the time-dependence of the expansion parameter $H(t)$. Therefore there exists a Goldstone boson $\pi$ associated with the breaking of such symmetry
\begin{align}
	t &\rightarrow t - \xi \nn \\
	\pi &\rightarrow \pi +\xi,
\end{align}
such that $\pi$ realizes time diffeomorphism non-linearly, but $U\equiv t+\pi$ transforms linearly as a scalar.
Demanding that the underlying theory is invariant under spacetime diffeomorphism,
the most general effective action for the Goldstone boson $\pi$ is just a derivative expansion in $U$. The resulting action, truncated for simplicity at one-derivative per field, is then given by~\cite{Cheung:2007st}
\begin{align}
S=\int \mathrm{d}^{4} x \sqrt{-g}\left[\frac{1}{2} \Mp^{2} R-\Mp^{2} \dot{H} g^{\mu \nu} \partial_{\mu}(t+\pi) \partial_{\nu}(t+\pi)-\Mp^{2}\left(3 H^{2}+\dot{H}\right)\right.\nonumber \\
\left.+\sum_{n} \frac{M_{n}(t+\pi)^{4}}{n !}\left(1-g^{\mu \nu} \partial_{\mu}(t+\pi) \partial_{\nu}(t+\pi)\right)^{n}+\cdots\right],
\label{eq:action_original}
\end{align}
where $M_{n}(t+\pi)$ are Wilson coefficients that are generally time dependent. We will make the additional assumption that the time variation is negligible, $\dot M^4_n  \ll H M^4_n$. This choice corresponds to the addition of a global symmetry, $\pi \to \pi +c$, and enforces that all the correlation functions from inflation are scale invariant. 

In the EFT of Inflation, terms with more than one-derivative per field can be described geometrically in terms of the extrinsic curvature of the time-slices in unitary-gauge (i.e.~the gauge where $\pi=0$). Here we are dropping higher derivatives per field for two reasons: (1) in unitary gauge, a term of the form $\delta g^{00}\delta K_{\mu}^{\mu}$ cannot change the speed of sound or contribute non-zero three point amplitude in the flat-space limit (2) in the decoupling limit, $ (\delta K_{\mu}^{\mu})^n$ only contribute higher derivative terms. When we take the soft limit of the amplitude, the contribution from these higher derivative terms are in general subdominant compared to the terms in the action above. We leave the full investigation with extrinsic curvature to the future.

\begin{figure}[t]
	\begin{center}
		\includegraphics[width=6in]{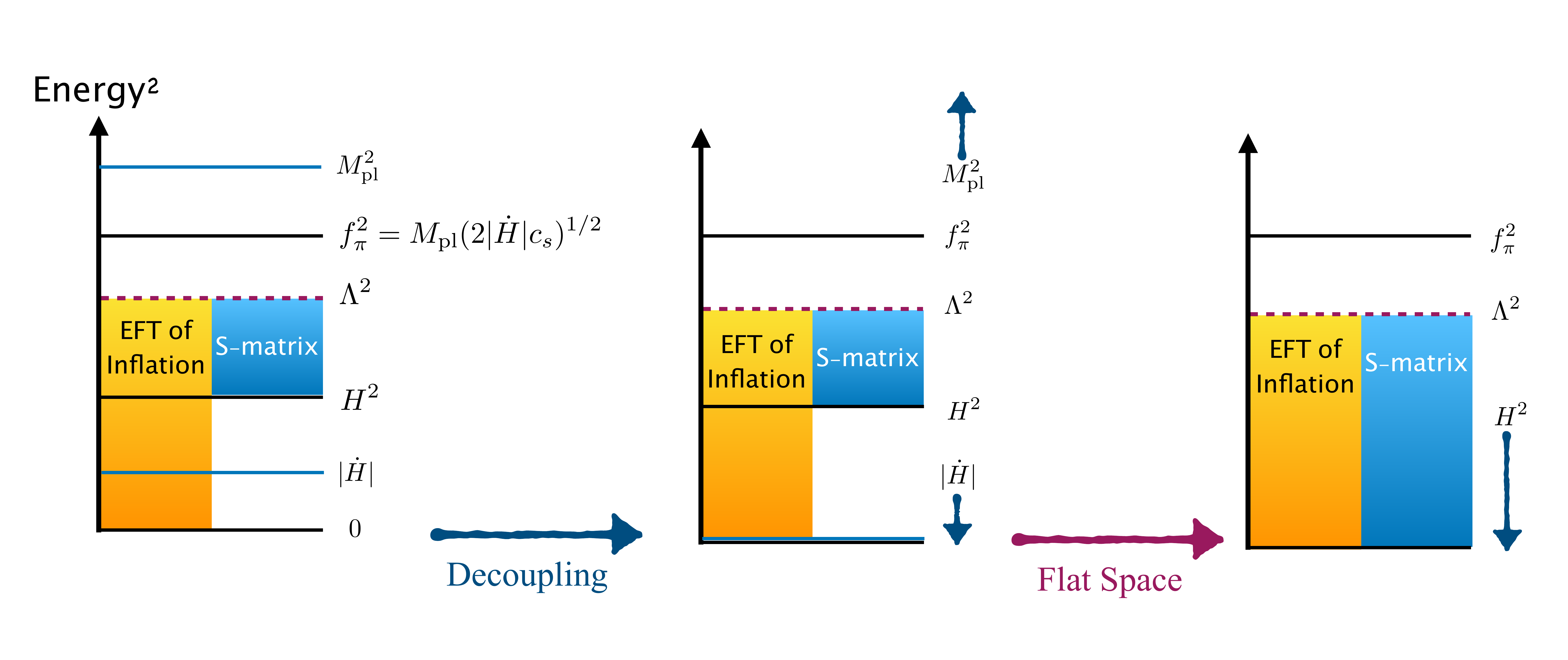}
	\end{center}
	\vskip -.5cm
	\caption{The hierarchy of scales that defines the EFT of Inflation. The scale $f_\pi^4 = 2\Mpl^2 |\dot H|c_s$ defines the Goldstone boson decay constant and is associated with the scale of symmetry breaking. The region in orange is where the EFT of Inflation in under control, while the region in blue is where flat-space scattering is a valid approximation. {\it Left:} The range of scales applicable to single-field inflation with a large non-Gaussian signature ($\Lambda < f_\pi$). {\it Center:} The decoupling limit around a fixed de Sitter background, $\dot H \to 0$, $\Mpl \to \infty$ with $f_\pi$ fixed.  {\it Right:} Flat-space limit of the EFT of Inflation, in the decoupling limit.  The soft theorems in the paper apply here where energies can be taken to zero.}
	\label{fig:Energy}
\end{figure}

We consider the sub-horizon limit of the EFT of Inflation in this paper, where spacetime reduces to flat background. The Goldstone boson propagates with a speed of sound $c_s \le 1$. Inflation is naturally described in terms of a hierarchy of (energy) scales~\cite{Baumann:2011su}
\begin{align}
 |\dot H|^2 \ll	H^4 \ll  f_\pi^4 \equiv 2 \Mp^2 |\dot{H}|c_s \ll \Mp^4
	\label{eq:ScaleHierachy}
\end{align}
where $f_\pi$ is the decay constant of Goldstone boson and the symmetry breaking scale, $f_\pi^4 = 2\Mpl^2 |\dot H| c_s$, illustrated in Figure~\ref{fig:Energy}. Concretely, in single-field slow-roll inflation ($c_s =1$), the scale of symmetry breaking is $f_\pi^4 = \dot \varphi^2$, which is the scale associated with the time evolution of the background scalar field $\varphi$. In our universe, $2 \Delta_{\zeta}^2 = H^4/f_\pi^4$, where $\Delta_{\zeta}^2$ is the amplitude of curvature perturbation, so that $f_\pi = 59 H$~\cite{Planck:2018jri}.  We will therefore consider energies $H^4 \ll E^4 \ll f_\pi^4$ so that the EFT of Inflation applies but we can neglect the curvature of spacetime. Within this context, we can consider scattering amplitudes for $\pi$ but they will not be Lorentz invariant.

The EFT of Inflation has an additional scale, $\Lambda = f_\pi c_s$ associated with the strong coupling, or a breakdown of the EFT description\footnote{It is conventional to write $M_n^4 = c_n f_\pi^4 (f_\pi/\Lambda)^{2n-1}$ with $c_n = {\cal O}((1-c_s^2))$ so $\Lambda$ controls the scale of irrelevant operators, after canonically normalizing $\pi$.}.  Weak coupling requires that $\Lambda > H$ but $\Lambda \ll f_\pi$ arises for $c_s \ll 1$. Validity of the EFT then requires $E < \Lambda$, which may further restrict the regime of validity of the flat-space approximation during inflation (in our universe). As our primary interest is to understand the structure of the EFT in general, we can take $H/E \to 0$ holding the other scales fixed so that the flat-space limit applies, as shown in Figure~\ref{fig:Energy}.
The soft limit we will use later should be taken \emph{after} this flat-space limit. In other words, the energy of the soft particle is small compared to others, but still much greater than $H$.

In the sub-horizon limit described above, the breaking of the time diffeomorphism reduces to the breaking of Lorentz boosts, but we still keep spacetime translation and $SO(3)$ spatial rotation.
The action in the flat-space limit reduces to
\begin{align}
	S=\int \mathrm{d}^{4} x \left[-\Mp^{2} \dot{H} g^{\mu \nu} \partial_{\mu}(t+\pi) \partial_{\nu}(t+\pi) 
	+\sum_{n} \frac{M_{n}^{4}}{n !}\left(1-g^{\mu \nu} \partial_{\mu}(t+\pi) \partial_{\nu}(t+\pi)\right)^{n}+\cdots\right].
	\label{eq:action_flat}
\end{align}
The metric can be set to $\eta^{\mu\nu}$ if we are not interested in graviton fluctuation\footnote{To make the flat-space limit precise, one often takes the decoupling limit, $\Mpl \to \infty$ and $\dot H \to 0$ holding $f_\pi$ fixed. One can relax this limit by considering large $\Mpl$ which allows (perturbative) graviton fluctuation.}.
Crucially, if the couplings are time-independent ($\dot M_n \to 0$), the Goldstone boson is derivatively coupled, reflecting the fact the that Goldstone has an additional (global) shift symmetry. As a consequence, the action (with no time dependent couplings) has an emergent time-translation symmetry that is distinct from the one that is generated by the stress-tensor.
As a result, we can still label the physical states by energy and momentum.
In the sub-horizon limit, the EFT of Inflation coincides with the $P(X)$ theory where the scalar obtains a vacuum expectation value that breaks Lorentz boosts
\begin{align}
	X = g^{\mu\nu}\partial_\mu (t+\pi) \partial_\nu (t+\pi)
	\label{eq:Xdef}
\end{align}
Although we obtain the action \Eq{eq:action_flat} by considering the sub-horizon limit of the EFT of Inflation, the results can be derived purely within flat spacetime by considering the breaking of Lorentz boosts while all other spacetime symmetries are preserved.
Therefore, the same action applies to other physical systems with the same symmetry breaking pattern. For instance, this action also describes the dynamics of superfluid~\cite{Son:2002zn}.

We can expand the action \Eq{eq:action_flat} in the order of the number of $\pi$-fields. We will make the unconventional choice of picking units such that
\begin{align}
	-2 \Mp^2 \dot{H} =1 \ .
\end{align}
Notice that with this choice $f_\pi^4 \to c_s$ and $\Lambda^4 \to c_s^5$.  The reason for this choice is to simplify the expression for our amplitudes while keeping track of the $c_s$ dependence. The non-linearly realized Lorentz boosts are defined in terms of the speed of light, $c$, and therefore we do not want to obscure the role of $c_s$ in their Ward identities.  In these units, the action reads
\begin{align}
	S &= \int d^4x\, \left[
	\frac{1}{2}+\dot{\pi} +\frac{1}{2}\left(
	c_s^{-2}\, \dot{\pi}^2 - (\partial_j \pi)^2
	\right)
	+ g_3 \dot{\pi}^3 + g_{3,1} \dot{\pi} \left(c_s^{-2}\, \dot{\pi}^2 - (\partial_j \pi)^2 \right)
	+\order\left(\pi^4\right)
	\right],
	\label{eq:action_new}
\end{align}
where $\dot{\pi}=\partial_0 \pi$ and $g_3$ and $g_{3,1}$ are coupling constants.
Matching to the original action in \Eq{eq:action_flat},
we find the deviation of $c_s$ from the speed of light is given by the coupling $M_2^4$ 
\begin{align}
	\delC \equiv \frac{1}{c_s^2}-1 = 4 M_2^4,
\end{align}
which is positive by causality.
The cubic couplings $g_3$ and $g_{3,1}$ are also related to $M_2^4$ and $M_3^4$
\begin{align}
	g_{3} &= -2 \delC M^4_2 - \frac{4}{3} M^4_3 \nn
		=-\frac{1}{2} \delC^2 - \frac{4}{3} M^4_3 \\
	g_{3, 1} &= 2 M^4_2 = \frac{1}{2}\delC \, .
\end{align}
The equation of motion (EOM) for $\pi$ up to quadratic order is then
\begin{align}
	\textrm{EOM} &= \partial_{\mu}\left(\frac{\delta \Lag}{\delta \partial_{\mu} \pi}\right)
	-\frac{\delta \Lag}{\delta \pi} \\
	&=\frac{1}{c_s^2}\ddot{\pi} - \nabla^2 \pi
	+3 g_{3} \partial_0 (\dot{\pi}^2)
	+\frac{\delC}{2} \partial_0 \left(
	\frac{3}{c_s^2} (\dot{\pi}^2)
	-(\partial_j \pi)^2
	\right)
	-\delC \,\partial_j (\dot{\pi} \partial_j \pi)
	+\order(\pi^3) \ .
	\label{eq:EOMpi}
\end{align}
Higher-orders terms in $\pi$ are straight-forward to determine from expanding Equation~(\ref{eq:action_flat}).

\subsection{Symmetry}
In the energy scale that we consider in \Eq{eq:ScaleHierachy} (with the assumption $\dot M_n = 0$), the breaking of time diffeomorphisms reduces to the breaking of Lorentz boosts. Consider an infinitesimal Lorentz boost along a spatial vector $\bm b$,
\begin{align}
	\delta x^\mu &= (\delta t, \delta \bm x) = (- \bm b \cdot \bm x, - \bm b \,t ) \\
	\delta \pi &= \bm b \cdot \left( \bm x +\bm x \dot{\pi} + t \bm{\nabla}\pi \right)
	= \bm b \cdot \bm x + \bm b \cdot \left(\bm x\, \partial_0 + t \bm{\nabla} \right)\,\pi \ ,
	\label{eq:symmetry_small}
\end{align}
where $\bm x\, \partial_0 + t \bm{\nabla}$ is the linear boost operator.\footnote{
In components, this is the usual boost as can be seen from $K^{i0}=x^i \partial^0 - t \partial^i = x^i \partial_0 + t \partial_i$.
} Crucially the transformation of $\pi$ includes the non-linear term, as well as the usual linear transformation. Note that the boost operator here is the standard relativistic boost, even when the Goldstone boson propagates with generic $c_s$.

The breaking of Lorentz boosts can be analyzed through conserved currents, which can be obtained via Noether approach or gravitational stress-energy tensor. We find the two approaches agree up to quadratic order which is sufficient to derive the soft theorem.
The stress-energy tensor can be obtained via
\begin{align}
	T_{\mu\nu} 
	&= 2 \frac{\delta \mathcal{L}}{\delta g^{\mu\nu}} - \eta_{\mu\nu} \mathcal{L} \\
	&= 2\partial_{\mu}(t+\pi)\partial_{\nu}(t+\pi) \frac{d\mathcal{L}}{dX} - \eta_{\mu\nu} \mathcal{L}.
\end{align}
Recall that $X$ is defined in \Eq{eq:Xdef}.
In terms of the components of $T^{\mu\nu}$, we find
\begin{align}
	T^{00} = 1 + (1+\delC)\dot{\pi}+\dots,\quad
	T^{j0} = -
	\partial_j \pi+\dots,\quad
	T^{ji} = \delta^{ij} \dot{\pi}+\dots
\end{align}
up to higher orders in $\pi$.
Even though the stress-energy tensor contains linear terms in $\pi$, this does \emph{not} imply that the translations are spontaneously broken. One has to check whether the vacuum expectation value of an operator is transformed under the temporal and spatial translations, whose generators are $H=\int d^3\bm x T^{00}$ and $P^i=\int d^3\bm x T^{0i}$. In our case, $U=t+\pi$ is such an operator as $\langle U \rangle \neq 0$ and $\langle [H,U] \rangle \neq 0$. Recall that there can still be an emergent time-translation symmetry of the EFT that it not generated by $T^{00}$, as explained below Equation~(\ref{eq:action_flat}).

From the stress-energy tensor, the currents associated with Lorentz generators are defined as
\begin{align}
	M^{\mu \rho \sigma} = x^\rho T^{\mu\sigma}-(\rho\leftrightarrow \sigma) \ .
\end{align}
The current associated with a boost along direction $i$ is given by
\begin{align}
	J^{\mu,i} = M^{\mu i 0} = x^i T^{\mu 0} - t T^{\mu i}.
\end{align}
Combining the expressions above we find the current in terms of the Goldstone field
\begin{align}
	J^{0,i} &= x^i + (1+\delC) x^i \dot{\pi} + t\partial_i \pi +\dots\\
	J^{j,i} &= -\delta^{ij} t - x^i \partial_j{\pi} - \delta^{ij} t \dot{\pi} +\dots
\end{align}
The matrix elements for boost currents with a single Goldstone state are given by
\begin{align}
    \langle \pi(q)| J^{0,i} |0\rangle 
    &= i e^{iq\cdot x} \left(\frac{1}{c_s^2} x^i q^0-t q^i \right) 
    \label{eq:JboostT}
    \\
    \langle \pi(q)| J^{j,i} |0\rangle
    &= i e^{iq\cdot x} \left( x^i q^j -t \delta^{ij}q^0 \right) \ ,
    \label{eq:JboostX}
\end{align}
which one can check that they obey the Goldstone theorem in \cite{Alberte:2020eil} and verify that the order parameters are non-zero in the broken phase.

Later we will use Ward identity to prove the corresponding soft theorem from spontaneously broken Lorentz boosts. Given a boost along $\bm b$, the current is
\begin{align}
	J^\mu = b^i J^{\mu,i},
\end{align}
and its divergence can be evaluated simply from EOM
\begin{align}\label{eq:NoetherEOM}
	\partial_{\mu} J^{\mu}
	&= \delta \pi \cdot \textrm{EOM} \\
	&= \bm b \cdot 
	\bm x \, (c_s^{-2} \ddot{\pi} - \partial_j^2 \pi)+\mathcal{O}(\pi^2) \nn
\end{align}
By construction, the current is conserved under the EOM. But the non-linear term in $\delta \pi$ leads to a nonzero matrix element between the vacuum and the one-particle state of the Goldstone boson.
This will be the starting point of the soft theorem derivation.

Of course all of the discussions so far are well-known. However, the use of local fields and conserved currents obscures the actual physical behavior. To elaborate on this point, consider the following change of field basis~\cite{Grall:2020ibl}
\begin{align}\label{eq:newPi}
	\pi \rightarrow \pi +\Delta \pi = \pi +\alpha \pi \dot{\pi}\,.
\end{align}
This induces a change on the action $\mathcal{L} \rightarrow \mathcal{L}+\delta \mathcal{L}$, where
\begin{align}
	\delta \mathcal{L} &=
	-\textrm{EOM}\cdot \Delta \pi \nn \\
	&=-(c_s^{-2}\ddot{\pi}-\nabla^2 \pi ) \cdot (\alpha \pi \dot{\pi}) +\order(\pi^2) =\frac{1}{2}\alpha \dot{\pi} (c_s^{-2}\dot{\pi}^2-(\nabla \pi)^2 )   +\order(\pi^2),
	\label{eq:action_newbasis}
\end{align}
where we use $\pi \dot{\pi}\Box \pi =-\frac{1}{2} \dot{\pi}(c_s^{-2}\dot{\pi}^2-(\nabla \pi)^2 )$ modulo total derivative. This implies a shift in the cubic coupling $g_{3,1} \rightarrow g_{3,1} + \alpha/2$ in the action \Eq{eq:action_new}. Given that $g_{3,1}=\delta_c/2$, we can eliminate this vertex by choosing
\begin{align}
	\alpha = -\delta_c \,.
	\label{eq:alphaValue}
\end{align}
The fact that the vertex $\dot{\pi} (c_s^{-2}\dot{\pi}^2-(\nabla \pi)^2 )$ can be removed by field definition is not surprising since the corresponding three-particle amplitude vanishes.

Crucially, the change of field basis also modifies the transformation of $\pi$ under non-linearly realized boosts. The transformation of the field now becomes
\begin{align}
	\delta \pi 
	= \bm b \cdot \bm x + \bm b \cdot \left(\bm x\, \partial_0 + t \bm{\nabla}\right)\pi 
	\rightarrow
	\delta \pi &= \bm b \cdot \bm x + \bm b \cdot \left(\bm x\, \partial_0 + t \bm{\nabla}\right)\pi 
	-\alpha \dot{\pi} \delta \pi
	+\order{(\pi^2)} \nn \\
	&=\bm b \cdot \bm x + \bm b \cdot \left((1-\alpha)\bm x\, \partial_0 + t \bm{\nabla} \right)\pi +\order{(\pi^2)},
	\label{eq:boost_generalBasis}
\end{align}
where terms that are higher order in $\pi$ under the new basis are omitted.
The crucial difference here is that the linear transformation now depends on $\alpha$.
In other words, we show that the symmetry transformation in \Eq{eq:symmetry_small} is not invariant under field redefinitions. Using the value in \Eq{eq:alphaValue} yields
\begin{align}
	\delta \pi 	&= \bm b \cdot \bm x + \bm b \cdot \left(\frac{1}{c_s^2}\bm x\, \partial_0 + t \bm{\nabla} \right)\pi  +\order{(\pi^2)}
	\label{eq:newTransformation}
\end{align}
where we define the boost operator with generic $c_s$ as
\begin{align}
	\frac{1}{c_s^2}\bm x\, \partial_0 + t \bm{\nabla} \,.
	\label{eq:newBoostPos}
\end{align}
In particular, this operator commutes with lightcone defined using speed $c_s$
\begin{align}
	\left(\frac{1}{c_s^2}\bm x\, \partial_0 + t \bm{\nabla}\right)\cdot \left(c_s^2 t^2 -\bm x^2 \right) = 0.
\end{align}
In momentum space, this implies boosts that preserve the on-shell condition are the ones with nontrivial $c_s$. As we will see, these are the correct boost operators used in the soft theorem with generic $c_s$.

\subsection{Amplitudes}
We use all out-going convention. For a scalar with outgoing momentum $p^\mu = (E,\bm p)$, the corresponding Feynman rule for $\partial_\mu \phi$ then yields $ip_\mu = (iE, -i \bm p)$.
Since boost invariance is spontaneously broken, the n-particle scattering amplitudes no longer only depend on the Lorentz-invariant momentum inner product. For Goldstone boson scattering, we can always write the amplitude $\amp_n$ as a function of the energy $E_i$ and rescaled inner product
\beq
\pp_{ij} \equiv 2\tilde{p}_i\cdot \tilde{p}_j \equiv c_s^{-2}E_iE_j-\bm p_i\cdot \bm p_j
\label{eq:p_def}
\eeq
between external momenta $p_i$ and $p_j$. The on-shell condition then reads
 \begin{align}
 	\tilde{p}_i^2 =0 \ .
 \end{align}
Notice that we organized the action in Equation~(\ref{eq:action_new}) so that contraction of indices will given this rescaled inner product.

Three-particle amplitudes in a Lorentz-invariant theory is trivial for scalar, since all the Mandelstam variables vanish. However, since boosts are broken here, we find nontrivial three-particle amplitudes for the Goldstone boson
\beq
\amp_3= -6ig_{3}  E_{1} E_{2} E_{3} 
\label{eq:Apion_3}
\eeq
which only depends on energies. Note that $g_{3,1}$ does not appear, which is consistent with the earlier discussion that it can be removed by a field redefinition. See Ref.~\cite{Pajer:2020wnj} for a more general classification.

The four-particle amplitude is given by 
\begin{align}
\amp_4=\amp_{4,\text{contact}}+\amp_s+\amp_t+\amp_u
\end{align}
where $A_{s,\text{contact}}$ is a purely local term that comes from the four-particle vertex in the original action, and $\amp_s$, $\amp_t$, $\amp_u$ are the terms from $s,t,u$ exchange diagrams. The contact term reads
\begin{align}
\amp_{4,\text{contact}}=&24 \left(\frac{1}{8} \delC^3 + 2 \delC M^4_3 + \frac{2}{3} M^4_4\right) E_1E_2E_3E_4+ +\delC\left[\pp_{12}^{\,2}+\pp_{13}^{\,2}+\pp_{14}^{\,2}\right] \nonumber\\
&4 \left(-\frac{1}{4}\delC^2 - 2 M^4_3\right) \left[(E_1E_2+E_3E_4) \pp_{12}+(E_1E_3+E_2E_4) \pp_{13}+(E_1E_4+E_2E_3) \pp_{14}\right].\nn
\end{align}
Note that we have a new coupling $M_4^4$ entering the four-particle scattering.
The contributions from exchange diagrams read
\begin{align}
    \amp_s &=E_{12}^2\left[-18g_{3}^2\frac{E_1E_2E_3E_4}{\pp_{12}}-12g_{3}g_{3,1}(E_1E_2+E_3E_4)-8g_{3,1}^2 \pp_{12} \right]\,,\nn \\
    \amp_t &=\amp_s\vert_{(1,2,3,4)\rightarrow (1,4,2,3)}\,, \quad 
    \amp_u =\amp_s\vert_{(1,2,3,4)\rightarrow (1,3,2,4)}.
\end{align}
Note even though the amplitudes do not depend on the field basis we use, the couplings constants $g_{3,1} $ do change under our field redefinition. Here we have written the three and four-point amplitudes in terms of the original field basis.

As we can see from the action and the above explicit example, there is always one more new coupling constant $M^4_{n+1}$ in $\amp_{n+1}$ when comparing to $\amp_{n}$. Therefore, the soft limit of $\amp_{n+1}$ cannot be trivially related to $\amp_{n}$. As we will see, the Ward identity arranges the soft limit in a clever way to circumvent the mismatch.

\section{Soft Theorems}
\label{sec:softthm}
In this section, we will derive the soft theorems for scattering amplitudes associated with the spontaneously broken boosts in the EFT of Inflation. We begin with the Ward-Takahashi identity
\begin{align}
	\partial_{\mu} \langle 0 | T( J^{\mu}(x) \pi(x_1) \dots \pi(x_n)) |0\rangle =-i \sum^n_{a=1} \delta(x-x_a) 
	\langle 0 | T(\pi(x_1) \dots \delta \pi(x_a) \dots \pi(x_n)) |0\rangle \ .
	\label{eq:Ward_raw}
\end{align}
In order to apply this to amplitudes, we use the Lehmann-Symanzik-Zimmermann (LSZ) reduction to $\pi (x_i)$ and Fourier transform the current to momentum space
\begin{align}	
	\lim_{q\rightarrow 0}
	\int_x \,
	e^{iq \cdot x}\, 
	\langle p_1,\dots,p_n | \partial_{\mu} J^{\mu}(x) |0\rangle
	&=-
	\sum^n_{a=1} \lim_{q\rightarrow 0} \lim_{\tilde{p}_a^2\rightarrow 0}
	\tilde{p}_a^2\int_{x} \, e^{i(q+p_a)\cdot x} \, \langle \{p_i|i\neq a\} | \delta \pi(x) |0\rangle
	= 0\, ,
	\label{eq:Ward_mom}
\end{align}
where $\int_x\equiv \int d^4 x$ and we amputate $\pi(x_a)$ by applying $\lim_{\tilde{p}_a^2\rightarrow 0}\int_{x_a} e^{ip_a\cdot x} (-i\tilde{p}_a^2)$. The amputation leads to the one-particle state, $\langle p_i|$.
The subscript $i$ runs from $1$ to $n$ modulo the additional conditions we specified. 
Note that the momentum $q$ is injected into $\delta \pi(x)$ on the right-hand side (RHS). This causes a mismatch between the momentum $p_a$, the one we use for amputation, and the momentum $p_a+q$ associated with the operator insertion.
One needs to be careful with the order of imposing on-shell condition and taking the soft limit in intermediate steps.\footnote{
Since $\delta \pi(x)$ in the RHS generates a single-particle pole $1/(\tilde{p}_a+\tilde{q})^2$, the ratio $\tilde{p}_a^2/(\tilde{p}_a+\tilde{q})^2$ actually depends on the order of on-shell limit $p_a^2\rightarrow 0$ and the soft limit $q\rightarrow 0$.
}
Although the final conclusion is the same, we will impose the on-shell condition first before taking the soft limit for most of the discussion. The RHS of \Eq{eq:Ward_mom} vanishes in this case since the momentum of amputation differs from the momentum associated with the $\delta \pi(x)$.

The goal is to translate the Ward-Takahashi identity into statements about on-shell amplitudes, also known as Ward identity.
In particular, in the kinematic limit where the momentum $q$ becomes soft for both energy and spatial components
\begin{align}
	q^\mu = (\omega, \bm q) \rightarrow 0 \, .
\end{align}
To maintain momentum conservation, this can be achieved by tuning one of the hard momenta, which we pick to be $p_1$,
\begin{align}
	p_1 = -\sum_{i=2}^n p_i -q \, .
	\label{eq:p1prescription}
\end{align}
It is important to impose the on-shell condition on ${p}_1$, $c_1^{-2}E_1^2-\bm p_1^2 =0$, where we assume particle 1 has the speed of propagation $c_1$, which can differ from $c_s$ or $c$. Under momentum conservation the on-shell condition leads to
\begin{align}
	c_1^{-2} \left(\sum_{a=2}^n E_a\right)^2 - \left(\sum_{a=2}^n \bm p_a\right)^2
	=-2 \left(c_1^{-2} \omega \sum_{a=2}^n E_a-\bm q\cdot \left(\sum_{a=2}^n \bm p_a\right) \right)-(c_1^{-2}\omega^2 -\bm q^2) \, .
	\label{eq:p1Onshell}
\end{align}
This implies the left-hand side is actually $\order(q)$ under the soft limit which is needed for the soft theorem.

\subsection{Preliminary: semi-onshell currents}
\label{sec:bgCurrent}

\begin{figure}[t]
	\begin{center}
		\includegraphics[trim=0 0 6cm 15cm,clip,scale=.4]{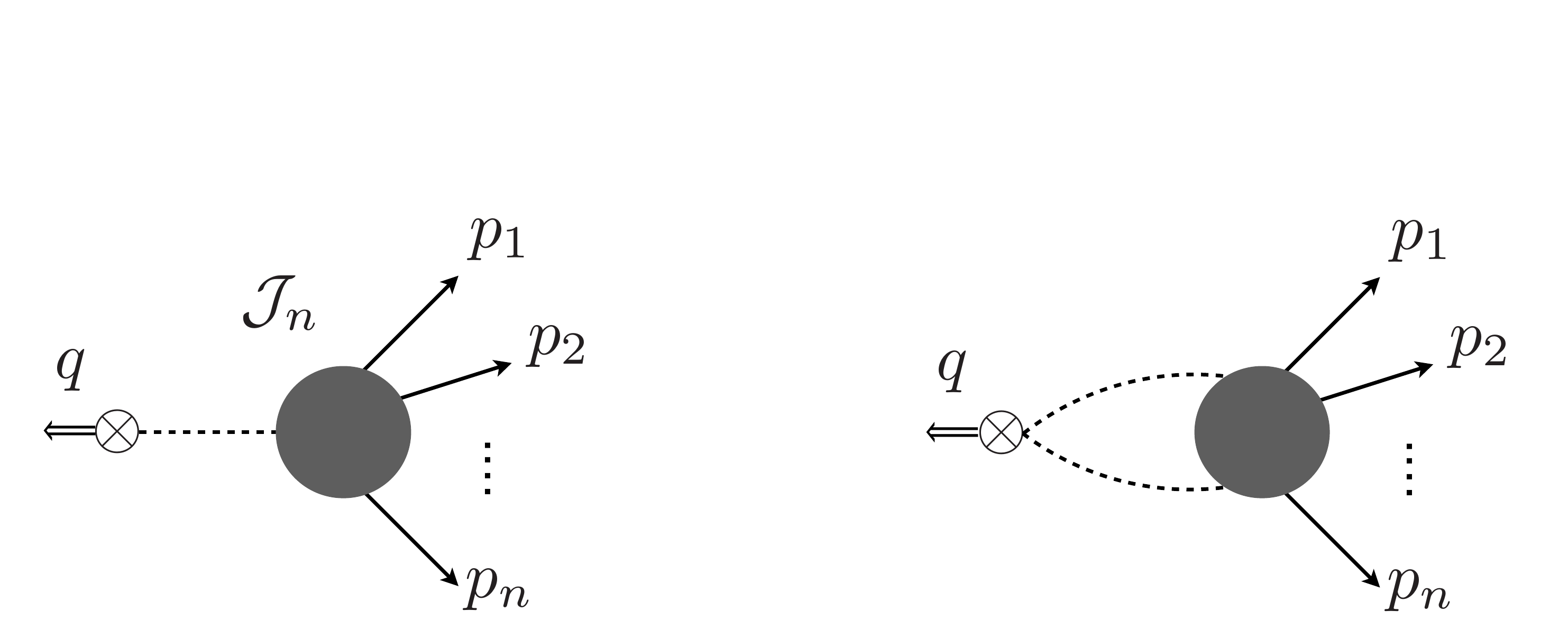}
	\end{center}
	\vskip -.5cm
	\caption{
		Left: the semi-onshell Berends-Giele current $\JBG_n$ that corresponds to $\pi(x)$ operator insertion between $n$-particle state and the vacuum. 
		Right: the matrix element of $\pi(x)^2$ operator insertion between $n$-particle state and the vacuum. 
		The solid external lines are on-shell particles with momentum $p_i$. The dashed lines denote  generically off-shell particles. When momentum $q$ becomes on shell, the current $\JBG_n$ on the left develops a $1/\tilde{q}^2$ singularity whose residue is given by the on-shell amplitude $\amp_{n+1}$.
	}
	\label{fig:BGcurrent}
\end{figure}

It is useful to consider the semi-onshell form factor of a field $\pi$ between an $n$-particle state and the vacuum, $\langle p_1,\dots,p_n |\pi(x) |0\rangle$, as a bridge between the off-shell correlation functions and on-shell amplitudes.
See Ref.~\cite{DiVecchia:2015jaq,Low:2017mlh,Low:2018acv,Cheung:2021yog} for previous applications to soft theorems.
This section reviews the basics of form factors and the closely related Berends-Giele (BG) current.
Readers familiar with the subject can skip this section.

The BG current is the Fourier transform of the form factor
\begin{align}
	\int_x \,
	e^{iq \cdot x}\, \langle p_1,\dots,p_n |\pi(x) |0\rangle 
	= \int_x \, e^{i (q+P)\cdot x} \, \JBG_n
	= \hat{\delta}(q+P)\, \JBG_n
	\label{eq:FF1}
\end{align}
where $\int_x \equiv \int \mathrm{d}^4x$, $P^\mu \equiv \sum_{i=1}^n p^\mu_i$ is the total momentum of external particles, and
$\JBG_n$ is the BG current with outgoing momentum $q$ induced by an $n$-particle state. 
See the left of \Fig{fig:BGcurrent} for the corresponding diagram.
We use the definition $\hat{\delta}(z) \equiv (2\pi)^4 \delta(z)$ throughout the paper.
In the first equality of \Eq{eq:FF1}, we factor out the $x$ dependence and $\JBG_n$ is defined as the rest of contribution.
In the on-shell limit, $\tilde{q}^2 \rightarrow 0$, the current develops a pole whose residue is given by the on-shell amplitude $\amp_{n+1}$ of the $n$ particles and the additional leg with momentum $q$
\begin{align}
	\JBG_n \xrightarrow{\tilde{q}^2\rightarrow 0} \frac{i}{\tilde{q}^{2}} \,i\amp_{n+1}
	+\dots,
	\label{eq:pole}
\end{align}
while the regular terms in the eclipses depends on off-shell degrees of freedom.
A special case is the overlap with a single-particle state
\begin{align}
	\langle p |\pi(x) |0\rangle 
	= e^{i p\cdot x},
	\label{eq:J1}
\end{align}
which implies $\JBG_1=1$.

\begin{figure}
	\begin{center}
		\includegraphics[trim=0 0 2cm 15cm,clip,scale=.35]{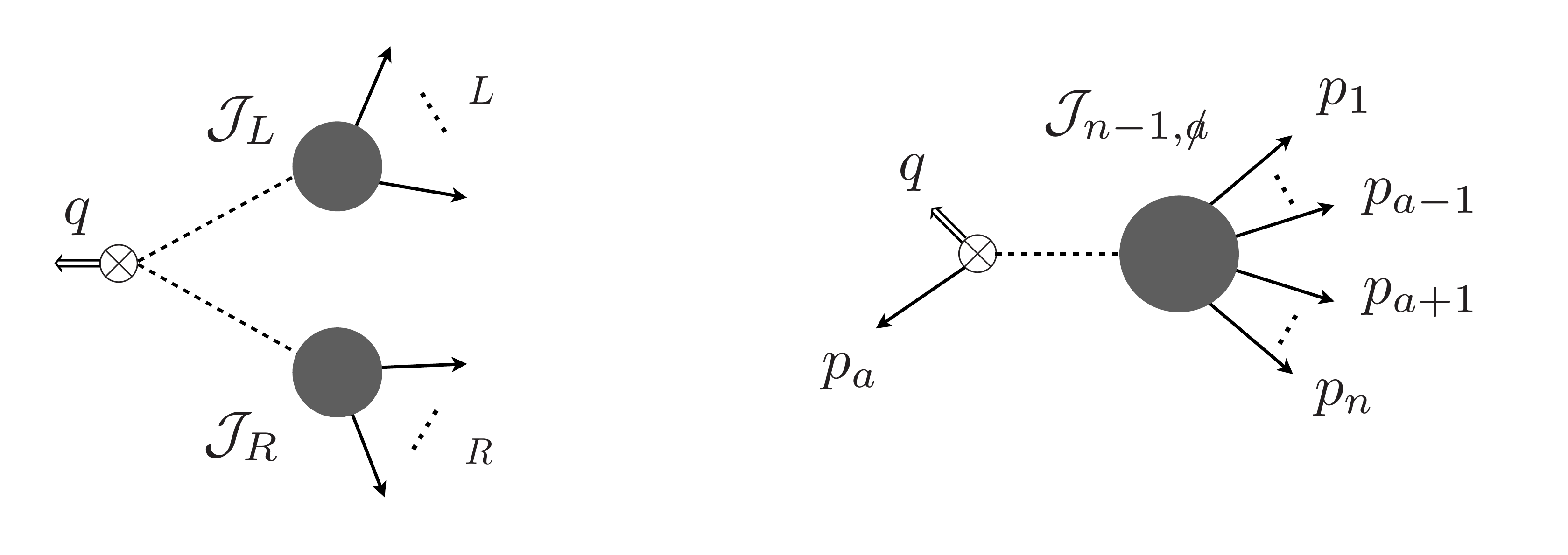}
	\end{center}
	\vskip -.5cm
	\caption{
		Left: the tree-level approximation of the matrix element of $\pi(x)^2$ shown in the right of \Fig{fig:BGcurrent}.
		We need to sum over all possible partitions $L$ and $R$ of on-shell particles.
		Right: the dominant subset of the same matrix element under the soft limit $q\rightarrow 0$, given by when either $L$ or $R$ contains only a single on-shell particle. The propagator of the dashed line scales as $\order(q^{-1})$ under the soft limit.
	}
	\label{fig:pi2Tree}
\end{figure}

We also need the form factor of $\pi(x)^2$ operator insertion, depicted in the right of \Fig{fig:BGcurrent}. 
The form factor can be evaluated in terms of BG current using perturbation theory. Using the tree-level approximation, the form factor is given by the sum over disconnected terms, as shown in the left of \Fig{fig:pi2Tree}, 
\begin{align}
	\int_x \,
	e^{iq \cdot x}\,\langle p_1,\dots,p_n |\pi(x)^2 |0\rangle 
	\xrightarrow{\rm tree} &\, 
	\int_x \,
	e^{iq \cdot x}\,\sum_{L,R}\langle \{p_i|i\in L\} |\pi(x) |0\rangle 
	\langle \{p_i|i\in R\} |\pi(x) |0\rangle, \nn \\
	=&\, \hat{\delta}(q+P)\, \sum_{L,R} \JBG_{L} \JBG_{R}
	\label{eq:FF2_Tree}
\end{align}
where we sum over all possible partitions of the $n$ particles into disjoint sets $L$ and $R$ whose corresponding BG currents are $\JBG_{L}$ and $\JBG_{R}$.
When we consider the soft limit $q\rightarrow 0$,
\Eq{eq:FF2_Tree} is dominated by the subset in which an internal propagator goes on-shell.
This subset is depicted by the right diagram in \Fig{fig:pi2Tree},
where either $L$ or $R$ contains a single particle $a$.
In this case, the dashed line is nearly on-shell since the momentum injection from $q$ is small, as can be seen from its propagator
\begin{align}
	\frac{i}{\prop_a} 
	\equiv \frac{i}{c_a^{-2}(E_a+\omega)^2-(\bm p_a+\bm q)^2}
	&= \frac{i}{c_a^{-2}(2\omega E_a+\omega^2) - (2\bm q \cdot \bm p_a+ \bm q^2)}\sim \order (q^{-1})\,,
	\label{eq:soft_prop}
\end{align}
where particle $a$ propagates with the speed of sound $c_a$, such that $c_a^{-2} E_a^2 -\bm p_a^2 =0$. Therefore if we first impose the on-shell condition and then take the soft limit, the propagator scales as $\order (q^{-1})$.
Combining \Eq{eq:J1} and the on-shell approximation we find
\begin{align}
	&\int_x \,
	e^{iq \cdot x}\,\langle p_a |\pi(x) |0\rangle \, \langle p_1,\dots, p_{a-1},p_{a+1},\dots,p_n |\pi(x) |0\rangle \nn \\[2pt]
	=&\, \hat{\delta}(q+P)\, \JBG_1\,\JBG_{n-1,\slashed{a}} \nn \\
	=&\, \hat{\delta}(q+P)\, \frac{i}{\prop_a}\, i\amp_{n}(p_1,\dots, p_{a-1}, p_a+q, p_{a+1}, \dots, p_n) +\dots \nn \\
	=&\, \hat{\delta}(q+P)\, \frac{i}{\prop_a}\, i\left(1+ q^\mu \frac{\partial}{\partial p_a^\mu}\right)\amp_{n}(p_1,\dots, p_n) +\dots
	\label{eq:Jsoft}
\end{align}
where we add the subscript $\slashed{a}$ to the current $\JBG_{n-1}$ denoting that $p_a$ is not included. 
In the second equality, we use $\JBG_1=1$ and only keep the singular part of the current $\JBG_{n-1,\slashed{a}}$ and replace the residue with an on-shell amplitude similar to \Eq{eq:pole}.
As written explicitly, the momentum of the $a$-th leg in the amplitude $\amp_{n}$ is extended to $p_a+q$ which maintains the conservation of total momentum.
This is an off-shell extension but the deviation is proportional to the inverse propagator $\prop_a$, such that off-shell deviation does not leave any non-local term behaves as $\order(q)/\prop_a$. 
In other words, all the residual terms are not only overall $\order(1)$ under the soft limit but also local in $q$.
Such an off-shell extension will be crucial for the subleading soft theorem.
In the last equality, we realize the extension by a differential operator on the hard amplitude.
Summing over $a$ on \Eq{eq:Jsoft} and including the symmetry factor, we finally arrive at the leading contribution to \Eq{eq:FF2_Tree} under the soft limit
\begin{align}
	\int_x \,
	e^{iqx}\,\langle p_1,\dots,p_n |\pi(x)^2 |0\rangle
	\xrightarrow{\textrm{tree},\,q\rightarrow 0} 
	&\, \hat{\delta}(q+P)\, \left[\sum_{a=1}^n \frac{2i}{\prop_a}\,\left(1+ q^\mu \frac{\partial}{\partial p_a^\mu}\right) i\amp_{n} +\order(1) \right].
	\label{eq:FF2_Simp}
\end{align}
We drop the momentum arguments of $\amp_{n}$ for simplicity.

\subsection{Goldstone boson amplitudes}
\subsubsection*{Relativistic Goldstone bosons}
Let us first consider the simple case with only Goldstone boson scattering and $c_s=1$, or equivalently $\delC=0$. In this case, we can use $\tilde{p}_i \cdot \tilde{p}_j = {p}_i \cdot {p}_j$ and $\prop_a=2p_a\cdot q$.
We are only interested in terms in the Ward-Takahashi identity up to $\order(1)$ under the soft limit.
Consider the left-hand-side (LHS) of \Eq{eq:Ward_mom}.
Even though the conserved current contains an infinite tower of terms, 
it drastically simplifies under the soft limit.
Since $\partial_{\mu}J^\mu(x) \sim q_\mu J^\mu(q)$ in momentum space,
most of their contribution to the Ward identity are only of the order of $\order(q)$.\footnote{Astute readers may be concerned here. Since $J^\mu(x)$ depends on the coordinate $x^\nu$, $\partial_{\mu} J^\mu(x)$ may contain terms that have no derivative and not suppressed by the soft limit.
For instance, $\partial_{\mu}(x^\nu f(x))= \delta^\nu_\mu f(x)+ x^\nu \partial_{\mu} f(x)$. However, the Fourier transform yields
\begin{align}
	\int_x e^{iq\cdot x} \partial_{\mu}(x^\nu f(x)) 
	=-q_\mu \left(\frac{\partial}{\partial q^\nu} f(q) \right),\nn
\end{align}
which is still suppressed by $q$ as long as $f(q)$ is regular in soft $q$.
}
The only possibility to get $\order(1)$ contribution is when the matrix element of $J^\mu(x)$ becomes singular. As we reviewed in the previous section, this only occurs in the form factors of $\pi(x)$ and $\pi(x)^2$,
whose singular behavior is given by \Eq{eq:pole} and \Eq{eq:FF2_Simp}.
This implies that we can truncate $\partial_{\mu} J^\mu(x)$ in \Eq{eq:Ward_mom} to quadratic order in the field.

To evaluate the matrix element of $\partial_{\mu} J^\mu(x)$, it is simpler to use Eq.~\eqref{eq:NoetherEOM} to relate it to the equation of motion~\Eq{eq:EOMpi}
\begin{align}
	\partial_{\mu} J^\mu(x) &= \delta \pi \cdot \textrm{EOM} \nn \\
	&= 
	\bm b \cdot \bm x\, \left(\Box \pi + 3 g_{3}\partial_0 (\dot{\pi})^2 \right)
	+ \bm b \cdot \left(\bm x\, \partial_0 + t \bm{\nabla}\right) \pi \, \Box \pi +\dots,
	\label{eq:current_relevant}
\end{align}
where we set $c_s=1$ ($\delC=0$) and truncate to quadratic order in $\pi$. The first and second parts originate from the non-linear and linear transformation of $\delta \pi$ interfering with the equation of motion.
All these terms can be evaluated using Eqs.~\eqref{eq:FF1}, \eqref{eq:pole} and \eqref{eq:FF2_Simp}. Let us discuss them in turn.

First, the contribution from the non-linear part of $\delta \pi$ leads to
\begin{align}
	&\lim_{q\rightarrow 0}\,\int_x e^{iq\cdot x} \,\bm b\cdot \bm x \langle p_1,\dots,p_n |\Box \pi + 3 g_{3}\partial_0 (\dot{\pi})^2 |0\rangle \nn \\
	=&\lim_{q\rightarrow 0}\, i\bm b \cdot \nabla_{\bm q}\,\left(\int_x e^{iq\cdot x} \,\langle p_1,\dots,p_n |\Box \pi + 3 g_{3}\partial_0 (\dot{\pi})^2 |0\rangle
	\right) \nn \\
	=&\lim_{q\rightarrow 0}\, i\bm b \cdot \nabla_{\bm q}\,
	\left[
	\hat{\delta}(q+P)\,\left(
	-q^2\JBG_{n}
	-\sum_{a=1}^n \amp_{3,a}\, 
	\JBG_{n-1,\slashed{a}}
	\right)
	\right],
	\label{eq:termNLRaw}
\end{align}
where $\nabla_{\bm q}$ is the derivative with respect to $\bm q$ and the vertex $3 g_3 \partial_0 (\dot{\pi})^2$ in momentum space reads
\begin{align}
	\amp_{3,a} \equiv \amp_{3}(q,p_a, -(q+p_a)) = 6i g_3 \omega E_a (\omega+E_a).
	\label{eq:v3_pi}
\end{align}
We denote this vertex by the same notation as the three-particle amplitude since they coincide. The three-particle amplitude is written in terms of energies which trivializes the off-shell extension of the leg with momentum $(q+p_a)$.

To express the above in terms of on-shell amplitudes, we use the dominant behaviors of $\JBG_{n}$ and $\JBG_{n-1,\slashed{a}}$ in \Eqs{eq:pole}{eq:FF2_Simp}.
Next, we only consider the term where the derivative $\nabla_{\bm q}$ acts on the parenthesis instead of the delta function $\hat{\delta}(q+P)$. This is well-defined since we already impose the choice in \Eq{eq:p1prescription} before taking the derivative which eliminate off-shell ambiguity from momentum conservation.\footnote{
\label{foonote:momConservation}
More rigorously, we can change the variable $p_1= \mathbb{P}-q-\sum_{i=2}^n p_i$, where $\mathbb{P} = P+q$ is the total momentum. Evaluating the derivative in \Eq{eq:termNLRaw} then yields the one acts on $\hat{\delta}(\mathbb{P})$ and the one acts on the parenthesis. Since they are formally separated when integrating with a test function, we only keep the latter.
}
Combining the above yields
\begin{align}
	&\lim_{q\rightarrow 0}\,\int_x e^{iq\cdot x} \,\bm b\cdot \bm x \langle p_1,\dots,p_n |\Box \pi + 3 g_{3}\partial_0 (\dot{\pi})^2 |0\rangle \nn \\	
	=&\hat{\delta}(q+P)\, i\bm b\cdot \nabla_{\bm q}\,
	\left(
	\amp_{n+1}
	+\sum_{a=1}^n \frac{\amp_{3,a}}{2p_a\cdot q}\, 
	\left(1+ q^\mu \frac{\partial}{\partial p_a^\mu}\right)
	\amp_{n}
	\right)
	+\order(q),
	\label{eq:termNL}
\end{align}
where we drop the momentum arguments in $\amp_{n+1}$ for simplicity.
In the presence of three-particle coupling, there is an $\order(q^{-1})$ soft singularity in $\amp_{n+1}$ as a result of factorization. 
But we observe that the $\amp_{n}$ term precisely cancels this singularity in $\amp_{n+1}$. This is not a coincidence, since this combination originates from the equation of motion which should vanish in a correlation function modulo local contact terms.
From \Eq{eq:termNLRaw} to \Eq{eq:termNL}, we replace $\JBG_{n}$ and $\JBG_{n-1,\slashed{a}}$ with amplitudes which are both valid up to $\order(1)$ correction. These corrections can only lead to the $\order(q)$ terms in the above equation which is the same order as other terms we drop in the current conservation.
While this is obvious to see for $\nabla_{\bm q}(q^2\JBG_{n})$ by naive counting, one needs to be careful with the $\order(1)$ residual terms in $\JBG_{n-1,\slashed{a}}$. The specific off-shell extension in \Eq{eq:Jsoft} is important here, since it ensures that the $\order(1)$ correction in $\JBG_{n-1,\slashed{a}}$ can only be local in $q$. The derivative $\nabla_{\bm q}$ on this local $\order(1)$ correction can only be at most $\order(1)$ under the soft limit.
Combining with the fact that $\amp_{3,a}$ only depends on $\omega$ but not $\bm q$, we find $\nabla_{\bm q} (\amp_{3,a} \times \order(1)) \sim \amp_{3,a} \times \nabla_{\bm q} (\order(1)) \sim \amp_{3,a} \times \order(1) \sim \order(q)$ if the $\order(1)$ term is local in $q$.
Therefore we conclude that the above equation is indeed valid.

The linear part of $\delta \pi$ can be evaluated similarly using tree-level expansion
\begin{align}
	&\lim_{q\rightarrow 0}\,\int_x \, e^{iq\cdot x} \, \langle p_1,\dots,p_n |(\bm b \cdot \left(\bm x\, \partial_0 + t \bm{\nabla} \right) \pi(x)) \, \Box \pi(x) |0\rangle \nn \\
	=&\lim_{q\rightarrow 0}\,\int_x \, e^{iq\cdot x} \, 
	\sum_{a=1}^n \Big[ \langle \{p_i|i\neq a\} |\bm b \cdot \left(\bm x\, \partial_0 + t \bm{\nabla} \right) \pi(x) |0\rangle 
	\langle p_a |\Box\pi(x) |0\rangle \nn \\
	&\qquad\qquad\qquad\quad +\langle \{p_i|i\neq a\} |\Box\pi(x) |0\rangle 
	\langle p_a | \bm b \cdot \left(\bm x\, \partial_0 + t \bm{\nabla} \right) \pi(x)|0\rangle \Big] \nn \\
	=&
	\sum_{a=1}^n \lim_{q\rightarrow 0} \lim_{p_a^2\rightarrow 0} 
	\left(
	-\bm b\cdot \Kmom_{a} \,\int_x  \, e^{i(p_a+q)\cdot x}
	\langle \{p_i|i\neq a\} |\Box\pi(x) |0\rangle 
	\right) \nn \\
	=&
	\sum_{a=1}^n \lim_{q\rightarrow 0} \lim_{p_a^2\rightarrow 0} 
	\bm b\cdot \Kmom_{a} \,\left[
	\hat{\delta}(q+P) \left((q+p_a)^2 \JBG_{n-1,\slashed{a}} \right)
	\right].
	\label{eq:termLinearRaw}
\end{align}
To arrive at the third line, we use the following identities on the single-particle state
\begin{align}
	\langle p_a |\Box\pi(x) |0\rangle  &= -p_a^2 e^{ip_a\cdot x} \xrightarrow{p_a^2\rightarrow 0} 0 \\[2pt]
	\langle p_a |\bm b \cdot \left(\bm x\, \partial_0 + t \bm{\nabla} \right) \pi(x) |0\rangle 
	&= \bm b\cdot \bm x \langle p_a |\dot{\pi}(x) |0\rangle
	+t \bm b\cdot \langle p_a |\bm \nabla{\pi}(x) |0\rangle \nn \\[3pt]
	&= -\bm b\cdot \Kmom_{a} \, e^{ip_a\cdot x},
	\label{eq:boostSingleParticle}
\end{align}
and integration by parts.
Here the relativstic boost generator $\Kmom_{a}$ for particle $a$ reads
\begin{align}
	\Kmom_{a} \rightarrow
	E_a\,\frac{\partial}{\partial \bm p_a}
	+\bm p_a \frac{\partial}{\partial E_a}.
\end{align}
Note that for generic $c_s$, the above will be replaced with a more general definition in \Eq{eq:boost_general}.
Similar to \Eq{eq:termNL},  we move the boost operator past the momentum-conserving delta function in \Eq{eq:termLinearRaw} and use \Eq{eq:FF2_Simp} to replace $\JBG_{n-1,\slashed{a}}$. This yields
\begin{align}
	&\lim_{q\rightarrow 0}\,\int_x \, e^{iq\cdot x} \, \langle p_1,\dots,p_n |(\bm b \cdot \left(\bm x\, \partial_0 + t \bm{\nabla} \right) \pi(x)) \, \Box \pi(x) |0\rangle \nn \\
	=&
	-\hat{\delta}(q+P)\,\sum_{a=1}^n
	\bm b\cdot \Kmom_{a} \, \left[\left(1+ q^\mu \frac{\partial}{\partial p_a^\mu}\right)\amp_{n}\right] +\order(q)\, \nn \\
	=&
	-\hat{\delta}(q+P)\,\sum_{a=1}^n
	\bm b\cdot \Kmom_{a} \, \amp_{n} +\order(q)\,.
	\label{eq:termLinear}
\end{align}
Unlike \Eq{eq:termNL}, the off-shell extension $p_a \rightarrow p_a+q$ is only $\order(q)$ so we can simply use the hard amplitude $\amp_n$ in the last line.

Combining \Eqs{eq:termNL}{eq:termLinear}, we find the Ward-Takahashi identity~\eqref{eq:Ward_mom} yields the soft theorem for spontaneously broken boosts on on-shell amplitudes
\begin{align}
	\lim_{q\rightarrow 0} i\nabla_{\bm q}\,
	\left(
	\amp_{n+1}
	+\sum_{a=1}^n \frac{\amp_{3,a}}{2p_a\cdot q}\, \left(1+ q^\mu \frac{\partial}{\partial p_a^\mu}\right)
	\amp_{n}
	\right)
	=&
	\sum_{a=1}^n \Kmom_{a} \,
	\amp_{n}
	+\order(q).
	\label{eq:ward_final}
\end{align}
This relates the soft Goldstone boson emission, after subtracting out the singularity from three-particle amplitudes,
to the boost of hard amplitude.
We verify the above theorem with tree-level scattering amplitudes up to $n=7$ by explicit calculation.

Let us emphasize the caveats in evaluating the above soft theorem, since derivatives acting on on-shell amplitudes are not always well-defined.
For soft theorems beyond leading order, it is common to take certain prescriptions in order for the theorems to hold~\cite{DiVecchia:2015jaq,Low:2017mlh,Low:2018acv,Cheung:2021yog}. And our soft theorem here is no exception.
As explained below \Eq{eq:termLinear}, the off-shell extension in $\Kmom_{a} \,
\amp_{n}$ is not relevant under the soft limit. 
On the LHS, we avoid the ambiguities from momentum conservation and on-shell conditions by applying the prescription \Eqs{eq:p1prescription}{eq:p1Onshell} \emph{before} taking the derivative $\nabla_{\bm q}$.

As we pointed out earlier, since $\amp_{n+1}$ contains one more coupling constant $M^4_{n+1}$ compared to $\amp_n$, the soft limit of the former cannot be fully fixed by the latter. Here we see that the Ward identity only relates the \emph{spatial derivative} of the soft emission to the hard amplitude. Since the coupling $M^4_{n+1}$ only contributes to $\amp_{n+1}$ as a contact term that only depends on energies, this new coupling is projected out by the derivative and therefore does not obstruct the soft theorem. On the other hand, we don't have full control of the soft limit which is needed to construct on-shell recursion relations for amplitudes~\cite{Cheung:2015ota,Luo:2015tat,Bartsch:2022pyi}.

\subsubsection*{Goldstone bosons with generic $c_s$}
For the case of generic $c_s$, the form of the soft theorem is not immediately obvious, since the boost operator actually depends on the field basis, as we have demonstrated explicitly in \Eq{eq:boost_generalBasis}.
Scattering amplitudes, which are always defined up to on-shell conditions, provide a clean perspective here. In order to have a valid soft theorem, the boost operator has to commute with on-shell conditions (at least before imposing momentum conservation). This is only true if the boost is with respect to $c_s$.
We will see this is indeed the case.

Mathematically, the soft theorem for generic $c_s$ can be derived in any basis. 
However, it is much easier to use the basis in \Eq{eq:newPi} with $\alpha=-\delC$ since the current conservation is almost identical to the relativistic case
\begin{align}
	\partial_{\mu} J^\mu(x) &= \delta \pi \cdot \textrm{EOM} \nn \\
	&= 
	\bm b \cdot \bm x\, \left((c_s^{-2}\partial^2_0 -\bm \nabla^2) \pi + 3 g_{3}\partial_0 (\dot{\pi})^2 \right)
	+ (\bm b \cdot (c_s^{-2} \bm x \partial_0+ t\bm \nabla) \pi) \, (c_s^{-2}\partial^2_0 -\bm \nabla^2) \pi +\dots\,.
\end{align}
Comparing to the relativistic case in \Eq{eq:current_relevant}, we only have to modify the boost and d'Alembert operators with respect to nontrivial $c_s$.
But crucially we stick to the same three-particle amplitude $\amp_{3,a}$ in \Eq{eq:v3_pi}, since the additional cubic vertex when $\delC\neq 0$ is canceled in the new basis by setting $\alpha=-\delC$ in \Eq{eq:action_newbasis}.
Therefore we just need to change the soft theorem with the new propagators and boost operators.
The boost operator for particle $a$ with respect to the speed of propagation $c_a$ is given by
\begin{align}
	\Kmom_{a} \equiv \frac{E_a}{c_a^2}\,\frac{\partial}{\partial \bm p_a}
	+\bm p_a \frac{\partial}{\partial E_a} \,.
	\label{eq:boost_general}
\end{align}
For pure Golstone boson scattering we have $c_a=c_s$.
In particular, the boost operator commutes with the on-shell condition of particle $a$,
\begin{align}
	\Kmom_{a}\left(p_a^2 \right) = \Kmom_{a}\left(E_a^2/c_a^2 - \bm p_a^2\right) =0
\end{align}
The soft theorem with generic $c_s$ then has exactly the same form as \Eq{eq:ward_final} but with the full boost operator~\eqref{eq:boost_general} and propagator $\prop_a$ defined in \Eq{eq:soft_prop}
\begin{align}
	\lim_{q\rightarrow 0}\, i\nabla_{\bm q}\,
	\left(
	\amp_{n+1}
	+\sum_{a=1}^n \frac{\amp_{3,a}}{\prop_a}\, \left(1+ q^\mu \frac{\partial}{\partial p_a^\mu}\right)
	\amp_{n}
	\right)
	=&
	\sum_{a=1}^n \Kmom_{a} \,
	\amp_{n}
	+\order(q).
	\label{eq:ward_finalcs}
\end{align}
We also check this up to $n=7$. The same caveats on off-shell extensions, discussed in the paragraph after \Eq{eq:ward_final}, apply to here as well.

The same soft theorem can be derived in the original field basis, although the calculation is more involved. The relativistic boost operator needs extra care since it does not commute with the propagator.
In the end, the additional three-particle vertex modifies the relativstic boost operator into the one with respect to $c_s$.
We can see that the standard basis, while it yields a simple action, does obscure the actual infrared behavior of physical observables.

\subsection{Coupling to Matter}\label{sec:mixed}
During inflation, we can have spectator fields other than the inflaton and the metric. 
In general, the spectator can have its own speed of propagation that is neither $c$ or $c_s$. Let us first review the construction in terms of action, and then discuss the soft theorem.

Consider a scalar field $\phi$ that propagates with speed $\cm$ whose action is
\begin{align}
	\mathcal{L}_\phi \supset \frac{1}{2} \left(
	\cm^{-2}\, \dot{\phi}^2 - (\nabla\phi)^2
	\right) +\dots\,. \nn
\end{align}
If we want to realize Lorentz invariance non-linearly, the deviation from relativistic kinetic term has to be assisted by coupling to Goldstone boson
\begin{align}
	\mathcal{L}_\phi =& \frac{1}{2} \partial^\mu\phi \partial_{\mu}\phi +\frac{\delPhi}{2}\, \left(\partial^\mu (t+\pi) \partial_{\mu}\phi \right)^2 \\[3pt]
	&+\frac{y_1}{2}\,(1-g^{\mu\nu}\partial_{\mu}(t+\pi)\partial_{\nu}(t+\pi)) \partial^\rho \phi \partial_\rho \phi +\frac{y_2}{2}\,(1-g^{\mu\nu}\partial_{\mu}(t+\pi)\partial_{\nu}(t+\pi)) \left(\partial^\mu (t+\pi) \partial_{\mu}\phi \right)^2,\nn
	\label{eq:matterFullAction}
\end{align}
where we define $\delPhi \equiv \cm^{-2}-1$.
To illustrate the full soft theorem, 
we also include two possible interactions starting at cubic order with couplings $y_1$ an $y_2$.
In general there are other possible interactions one can write down.
The theory is now also invariant under boosts. Under the infinitesimal transformation in \Eq{eq:symmetry_small}, the scalar $\phi$ undergoes the linear relativistic boost
\begin{align}
	\delta \phi = 
	\bm b \cdot \left(\bm x\, \partial_0 + t \bm{\nabla} \right)\phi \,.
\end{align}

Crucially the non-linearly realized Lorentz invariance implies that the modification of kinetic term comes hand-in-hand with a cubic interaction with the Goldstone boson. Expanding \Eq{eq:matterFullAction} to cubic order yields
\begin{align}
	\mathcal{L}_\phi = \frac{1}{2} \left(
	\cm^{-2}\, \dot{\phi}^2 - (\nabla\phi)^2
	\right)
	+\delPhi\, \dot{\phi}\, \partial^\mu \phi \partial_\mu \pi -y_1 \dot{\pi} (\partial \phi)^2 -y_2 \dot{\pi} \dot{\phi}^2 +\dots
	\label{eq:matterCubicAction}
\end{align}
We can also compute the corresponding amplitudes. For instance, the amplitude with two $\phi$ and one $\pi$ from \Eq{eq:matterCubicAction} reads
\begin{align}
	\amp_{3}(q_\pi,p_{\phi},(-q-p)_\phi) =& -2i(\delPhi^2+2y_1\delPhi+y_2)\,\omega E(E+\omega)- i\left(\frac{\delta_\phi}{2}-y_1\right)(\delPhi-\delC)\omega^3\,,
\end{align}
where the subscripts on the momenta label the particle species, i.e., particles with momentum $p$ and $-q-p$ in the above correspond to $\phi$ field and particle with momentum $q$ is the Goldstone boson.

As in the Goldstone boson case, we can also do a field redefinition to change the linear boost and the cubic vertices. Observe that under integration by parts
\begin{align}
	\delta_\phi \dot{\phi}\, \partial^\mu \phi \partial_\mu \pi
	=& -\delPhi^2\,\dot{\pi} \,\dot{\phi}^2 
	+\frac{\delta_\phi}{4}\left(\delPhi-\delC\right)\, \dddot{\pi} \phi^2 \nn\\
	&-\delta_\phi \phi\dot{\phi} \,\left(c_s^{-2}\partial^2_0- \bm \nabla^2\right) \pi
	-\delta_\phi \left(\pi\dot{\phi}+\frac{1}{2}\phi\dot{\pi} \right)\left(\cm^{-2}\partial^2_0- \bm \nabla^2\right) \phi.
	\label{eq:3ptIdentity}
\end{align}
The second line is proportional to the leading equations of motion. Thus they can be removed by the following field redefinitions
\begin{align}
	\pi \rightarrow& \pi + \Delta\pi
	= \pi - \delPhi \phi \dot{\phi} \nn \\
	\phi \rightarrow& \phi + \Delta\phi
	= \phi - \delPhi \left(\pi\dot{\phi}+\frac{1}{2}\phi\dot{\pi} \right),
	\label{eq:newBasis_matter}
\end{align}
where the shift $\Delta\pi$ and $\Delta\phi$ is related to the terms proportional to equations of motion~\eqref{eq:3ptIdentity}.

The transformation of $\phi$ and $\pi$ under the Lorentz boost is also modified accordingly. Since the original $\phi$ transform linearly, $\delta\pi$ remains the same up to linear order in the new basis. However, the original $\delta\pi$ has a non-linear term which modifies the $\delta\phi$ at linear order
\begin{align}
	\delta\phi &=\bm b \cdot \left(\bm x\, \partial_0 + t \bm{\nabla} \right)\phi +\delPhi \left(\delta\pi\dot{\phi}+\frac{1}{2}\phi\,(\partial_0{\delta\pi}) \right) +\order(\phi^2,\pi\phi) \nn \\
	&=
	\bm b \cdot \left(c_\phi^{-2} \bm x\, \partial_0 + t \bm{\nabla} \right)\phi.
	\label{eq:newTransformation_matter}
\end{align}
We find that under this field basis, $\delta\phi$ is given by the boost with respect to its own speed $c_\phi$.
The action in the new basis reads
\begin{align}
	\mathcal{L}_\phi = \frac{1}{2} \left(
	\cm^{-2}\, \dot{\phi}^2 - (\nabla\phi)^2
	\right)
	-\delPhi^2\,\dot{\pi} \,\dot{\phi}^2 
	+\frac{\delta_\phi}{4}\left(\delPhi-\delC\right)\, \dddot{\pi} \phi^2
	-y_1 \dot{\pi} (\partial \phi)^2 -y_2 \dot{\pi} \dot{\phi}^2 +\dots
	\label{eq:matterCubicActionNew}
\end{align}
Similar to the earlier discussion, the derivation of soft theorem with coupling to matter will be more straightforward in this field basis.

The soft theorem in the general case with $\pi$ and the matter $\phi$ also follows similarly. We use the new basis combining \Eq{eq:newBasis_matter} and \eqref{eq:newPi} (with $\alpha = -\delC$) such that the linear boosts of $\pi$ and $\phi$ are with respect to the its own speed of propagation $c_s$ and $c_\phi$.
So the soft theorem is still given by \Eq{eq:ward_finalcs}, but with the generalization to include the $\pi-\phi-\phi$ cubic vertex and the boost with the speed of each particle. The general three-particle vertices are
\begin{align}
	\amp_{3,a} 
	&= \amp(q_\pi,p_{a,X},(-q-p_a)_X)=
	\begin{cases}
		\amp(q_\pi,p_{a,\pi},(-q-p_a)_\pi) & \textrm{if $a\in \pi$} \\[5pt]
		\amp(q_\pi,p_{a,\phi},(-q-p_a)_\phi) & \textrm{if $a\in \phi$}
	\end{cases}
\end{align}
where $X$ denotes the species of particle $a$ and the off-shell leg $-q-p_a$
\begin{align}
	\amp(q_\pi,p_{a,\pi},(-q-p_a)_\pi) &= 6i g_3 \,\omega E_a (\omega+E_a) \\
	\amp(q_\pi,p_{a,\phi},(-q-p_a)_\phi) &= -2i(\delPhi^2+2y_1\delPhi+y_2)\,\omega E_a(E_a+\omega)- i\left(\frac{\delta_\phi}{2}-y_1\right)(\delPhi-\delC)\omega^3.
\end{align}
Again, it turns out that the vertices in all the cases coincide with the three-particle amplitudes when written in terms of energy.~\footnote{
From the action in \Eq{eq:matterCubicActionNew}, the vertex  actually has an extra term proportional to $\omega \prop_a$ which only leads to $\order(q)$ correction in the soft theorem. Therefore effectively the $\pi-\phi-\phi$ vertex is still given by the three-particle amplitude $\amp_{3,a}$. This is the case in general as long as the three-particle vertex is proportional to $\omega$.
}
If we include other interactions between the matter and the Goldstone boson, $\amp(q_\pi,p_{a,\phi},(-q-p_a)_\phi)$ needs to be modified accordingly.

\subsection{Non-perturbative validity}
The beauty of symmetry is that the many consequences are valid even non-perturbatively.
For instance, both the Ward-Takahashi identity and the Goldstone theorems on the correlation functions hold for any couplings.
Although our derivation starts with Ward-Takahashi identity, tree-level expansion is needed to evaluate the correlation functions of $\pi(x)^2$ shown in \Eq{eq:FF2_Simp}. Therefore, the full soft theorem is only valid for tree-level scattering amplitudes.
Nevertheless, the need of perturbative expansion can be circumvented when three-particle amplitudes vanish. In this case, our soft theorem can be lifted to non-perturbative level.
This is analogous to the Adler zero for soft pion emission in QCD, which also holds non-perturbatively under the same condition.\footnote{The vanishing of three-pion amplitude in QCD is guaranteed from parity, while the pion-nucleon-nucleon amplitude is not zero. So Adler zero holds non-perturbatively for pion scattering, but breaks down in the presence of nucleon.}

Let us specify the assumptions we use in the non-perturbative regime. 
\begin{enumerate}
	\item Lorentz boosts are spontaneously broken but not translation.
	\item The Goldstone boson propagates with speed of light, $c_s=1$.
	\item The matrix elements of the boost current $J^{\mu,i}$ between the single-particle state and the vacuum are given by \Eqs{eq:JboostT}{eq:JboostX}.
	\item No three-particle amplitudes and the soft limit of amplitudes starts at $\order(q)$. These two statements are equivalent at tree level but we list them separately for the sake of generic case.
\end{enumerate}
Consider the Ward-Takahashi identity~\eqref{eq:Ward_raw} under these assumptions. As we mentioned, one needs to be careful with the order between soft limit $q\rightarrow 0$ and the on-shell limits of hard particles, $p_a^2 \rightarrow 0$. In the earlier sections with generic three-particle amplitudes, we take the on-shell limits first and then the soft $q$ limit. 
In the perturbative regime with zero three-particle amplitudes, one can take the order of limits in either way and still find the same conclusion.
Here we will show the derivation of soft theorem in the opposite order which is valid when three-particle amplitudes vanish. As we will see, this derivation does not need perturbative expansion and therefore holds more generally under the assumptions.
Taking $q\rightarrow 0$ limit first and apply LSZ reduction on the Ward-Takahashi identity yields
\begin{align}	
	[\textrm{LSZ}]\lim_{q\rightarrow 0} \int_x e^{i q\cdot x}
	\partial_{\mu} \langle 0 | T( J^\mu(x) \pi(x_1) \dots \pi(x_n)) |0\rangle
	=-
	\sum^n_{a=1} \lim_{p_a^2\rightarrow 0}
	p_a^2\int_{x_a} \, e^{i p_a\cdot x} \, \langle \{p_i|i\neq a\} | \delta \pi(x) |0\rangle \nn
\end{align}
where $[\textrm{LSZ}]\equiv \prod^n_{a=1} \lim_{p_a^2\rightarrow 0} (-i p_a^2)\int_{x_a} e^{ip_a\cdot x_a}$ and we have already applied $q\rightarrow 0$ in the RHS. Let us consider each side of the equation in turn.

On the RHS, the momentum from the Fourier transform matches the one used for amputation. So it no longer vanishes in contrast to the other order of limits taken in \Eq{eq:Ward_mom}. Given that $\delta\pi(x)$ in \Eq{eq:symmetry_small}, we see that the non-linear term does not generate a pole and thus drops out, so we only need to keep the linear part in $\delta\pi(x)$
\begin{align}
	&-\sum^n_{a=1} \lim_{p_a^2\rightarrow 0}
	p_a^2\int_{x_a} \, e^{i p_a\cdot x} \, \langle \{p_i|i\neq a\} | \delta \pi(x) |0\rangle \nn \\
	=&-\sum^n_{a=1} \lim_{p_a^2\rightarrow 0}
	p_a^2\int_{x_a} \, e^{i p_a\cdot x} \, \langle \{p_i|i\neq a\} | \bm b\cdot (\bm x \partial_0+t\bm \nabla)\pi(x) |0\rangle \nn \\
	=&-\sum^n_{a=1} \lim_{p_a^2\rightarrow 0}
	p_a^2  (\bm b \cdot \bm \Kmom_a)\, \int_{x_a} \, e^{i p_a\cdot x} \, \langle \{p_i|i\neq a\} |\pi(x) |0\rangle \nn \\
	=&
	\sum^n_{a=1} 
	\bm b \cdot \Kmom_a\,\left(
	\delta(P) \amp_{n}
	\right)
	\label{eq:NP_rhs}
\end{align}
We use the fact that $p_a^2$ commutes with the boost operator $\Kmom_a$ with $c_s=1$.

The LHS is very similar to the tree-level calculation. The crucial difference is that we first take the soft limit before the amputation. 
\begin{align}	
	&[\textrm{LSZ}]\lim_{q\rightarrow 0} \int_x e^{i q\cdot x}
	\partial_{\mu} \langle 0 | T( J^\mu(x) \pi(x_1) \dots \pi(x_n)) |0\rangle \nn \\
	=&[\textrm{LSZ}] \lim_{q\rightarrow 0} \left[
	\int_x e^{i q\cdot x} x^i\,
	\langle 0 | T( \Box \pi(x) \pi(x_1) \dots \pi(x_n)) |0\rangle 
	+\order(q) \right]
\end{align}
We use the assumption that the only pole created by the current in an off-shell correlation function is the single-particle emission. So effectively we can use the leading term in $\partial_{\mu} J^{\mu,i} \sim x^i \Box \pi$. The rest of the contribution is suppressed by $\order(q)$ due to the derivative on the current.
It is crucial that we take the soft limit before on-shell limits. If this is not the case, the current can be inserted to an external on-shell line and generate a $1/(p_a+q)^2 \sim 1/(2p_a\cdot q)$ pole.\footnote{In the previous perturbative derivation, $\partial_{\mu}J^\mu$ contains quadratic terms in field, even in the absence of three-particle amplitudes, which leads to the boost on hard amplitude as shown in \Eqs{eq:termLinearRaw}{eq:termLinear}. But if we first take the soft limit, the quadratic in $\pi$ contribution is suppressed.
The soft theorem remains the same, but the boost on hard amplitudes is now reproduced by the other side of Ward identity as shown in \Eq{eq:NP_rhs}.}
Now keeping only the leading order contribution in the above equation leads to
\begin{align}
	[\textrm{LSZ}]\lim_{q\rightarrow 0} \int_x e^{i q\cdot x}
	\partial_{\mu} \langle 0 | T( J^\mu(x) \pi(x_1) \dots \pi(x_n)) |0\rangle
	=&[\textrm{LSZ}] \lim_{q\rightarrow 0} \bm b \cdot \nabla_{\bm q}
	\langle q | T( \pi(x_1) \dots \pi(x_n)) |0\rangle \nn \\
	=&\lim_{q\rightarrow 0} \bm b \cdot \nabla_{\bm q}
	\left( \delta(q+P) i\amp_{n+1}\right).
	\label{eq:NP_lhs}
\end{align}
When we commute the on-shell conditions with the derivative, one needs to apply the same prescription as before.

Equating both sides of the Ward identity, given in \Eq{eq:NP_rhs} and \Eq{eq:NP_lhs}, and factoring out the momentum conservation delta functions, we find
\begin{align}
	\lim_{q\rightarrow 0}\,i \nabla_{\bm q}\,
	\amp_{n+1} = \sum^n_{a=1} 
	\Kmom_a\,\amp_{n} +\order(q).
\end{align}
Crucially, we do not need perturbative expansion in this derivation,
under the assumptions listed earlier, therefore we believe it also holds non-perturbatively.
The final theorem is the same as the tree-level results~\eqref{eq:ward_finalcs} when $\mathcal{V}_{3,a}=0$ and $c_s=1$.
It is possible that the non-perturbative theorem also extends to generic $c_s$ when three-particle amplitudes vanish.
But we leave this investigation to the future.

\subsection{Summary}
\label{sec:softthm_summary}

In this section we derived a soft theorem for the spontaneous breaking of boosts in the EFT of Inflation.  The final form of this soft theorem is given by
\begin{center}
\begin{tcolorbox}[colback=light-gray]
\begin{minipage}{\textwidth}
\begin{align}
	& \lim_{q\rightarrow 0}\, i\nabla_{\bm q}\,
	\left(
	\amp_{n+1}
	+\sum_{a=1}^n \frac{\amp_{3,a}}{\prop_a}\, \left(1+ q^\mu \frac{\partial}{\partial p_a^\mu}\right)
	\amp_{n}
	\right)
	=
	\sum_{a=1}^n \Kmom_{a} \,
	\amp_{n}
	+\order(q).	
	\label{eq:summary} \\[5pt]
	&	
	\prop_a \equiv c_a^{-2}(E_a+\omega)^2-(\bm p_a+\bm q)^2 \,,\quad
	\Kmom_{a} \equiv  \frac{E_a}{c_a^2}\,\frac{\partial}{\partial \bm p_a}
	+\bm p_a \frac{\partial}{\partial E_a} \\
	&
	\amp_{3,a} 
	=
	\begin{cases}
		6i g_3 \,\omega E_a (\omega+E_a) & \textrm{if $a\in \pi$} \\[5pt]
		-2i(\delPhi^2+2y_1\delPhi+y_2)\,\omega E_a(E_a+\omega)- i\left(\frac{\delta_\phi}{2}-y_1\right)(\delPhi-\delC)\omega^3 & \textrm{if $a\in \phi$}
	\end{cases}
\end{align}
\end{minipage}
\end{tcolorbox}
\end{center}
The prescriptions in \Eqs{eq:p1prescription}{eq:p1Onshell} need to be applied \emph{before} taking the derivative $\nabla_{\bm q}$.
For matter coupling we assume the interaction in \Eq{eq:matterFullAction}.
When the three-particle amplitudes vanish and $c_a=1$ for all particles, the soft theorem becomes non-perturbative.
A non-trivial feature of this result, is the appearance of a modified boost operator $\Kmom_{a}$, associated with the speed of propagation of a given particle, $c_a$, even though it is derived from the Ward identity for the broken relativistic boosts (i.e.~$c=1$) of the microscopic theory.

\section{Implications for Inflation}
\label{sec:inflation}

\subsection{Dirac-Born-Infeld Inflation}
Our soft theorem applies to generic models of inflation. But we can ask if there are any ``special'' models for inflation from this on-shell point of view. For Lorentz-invariant theories, this is analogous to the ``exceptional'' Goldstone boson EFTs identified through the enhanced soft limit~\cite{Cheung:2014dqa,Cheung:2016drk}. As an example, the DBI action
\begin{align}
    S = \int d^4x \sqrt{1+\partial^\mu \phi \partial_\mu \phi }
    \label{eq:DBIAction}
\end{align}
is a special case of $P(X)$ theory, where there is one derivative per scalar field. 
When the momentum $q$ of a scalar becomes soft, we expect the amplitudes scale as $\mathcal{O}(q)$ which is the case for a generic $P(X)$ theory. However, the DBI model is special since the leading $\mathcal{O}(q)$ soft limit of  amplitudes cancels and the soft behavior is lifted to $\mathcal{O}(q^2)$. 
In fact, one can prove that the DBI theory is the unique one in the class of $P(X)$ models that has such a soft limit~\cite{Cheung:2014dqa,Cheung:2016drk}. 
The special soft limit can be linked to the symmetry in the DBI action
\begin{align}
    \delta_{\nu} \phi = x^\nu + \phi \partial^\nu \phi.
    \label{eq:DBIsymmetry}
\end{align}
This symmetry is natural when one views the DBI theory as the EFT of a $d$-brane in $d+1$ dimension, where $\phi(x)$ parameterizes the position of the brane in extra dimension~\cite{Leigh:1989jq}. The above symmetry is then the boost invariance involving the extra dimension. We can see that examining the soft limit is a bottom-up approach to identify special theories without diving to the details  in traditional symmetry construction.

Equipped with the universal soft theorem for the EFT of inflation in our disposal, we can ask if there is any special models of inflation from this point of view.
Back to the DBI theory, it is also a popular model for inflation due to its connection to string theory~\cite{Silverstein:2003hf,Alishahiha:2004eh} (see e.g.~\cite{Baumann:2014nda} for review). For the  applications to inflation, we consider the fluctuation around the boost-breaking vacuum
\begin{align}
    \phi = \langle \phi \rangle + \pi = \mu t + \pi.
\end{align}
However, as pointed out by Ref.~\cite{Grall:2020ibl}, one can use the symmetry in \Eq{eq:DBIsymmetry} to restore $\langle \phi \rangle$ to zero. Beyond the level of vacuum expectation value, Ref.~\cite{Grall:2020ibl} shows one can construct an unbroken boost from a linear combination of the original boost and the one from  \Eq{eq:DBIsymmetry}. The new symmetry generators form a Poincare algebra if one rescales time $t \rightarrow t/c_s$. In other words, the DBI theory in the boost-breaking vacuum has an emergent Lorentz invariance under the rescaled time. 
Motivated by this behavior, we ask the following question: \emph{``what are the boost-breaking EFTs that have an emergent Lorentz invariance with respect to the speed of sound?''}

We will see that DBI Inflation is the unique boost-breaking EFT that has such an emergent Lorentz invariance. The proof is quite simple.
After we rescale the time, the theory should behave like an ordinary Lorentz-invariant theory.
Therefore, the action for a derivatively-coupled $\pi$ must be in the form of a $P(X)$ theory.
This enforces the three-particle vertex to vanish.
In addition, the amplitudes are functions of the rescaled inner product, $\tilde{p}_i\cdot \tilde{p}_j$, defined in \Eq{eq:p_def}, and they satisfy $\sum_{a=1}^n \Kmom_{a} \,	\amp_{n}=0$.
On the other hand, the amplitudes must satisfy the soft theorem from the non-linearly realized boost.
Given the vanishing of three-particle vertices and $\sum_{a=1}^n \Kmom_{a} \,	\amp_{n}=0$, our soft theorem for an EFT with emergent Lorentz-invariance reduces to
\begin{align}
	&\lim_{q\rightarrow 0}\, i\nabla_{\bm q}\,
	\amp_{n+1}
	= 0
	\label{eq:DBI}
\end{align}
up to higher order terms in $q$. Since $\amp_{n+1}$ is a function of $\tilde{p}_i\cdot \tilde{p}_j$, the above soft theorem implies that the full $\order{(q)}$ soft limit of $\amp_{n+1}$ vanishes.
So after we rescale the time, the soft theorem from the boost demands that the Goldstone boson must be a $P(X)$ theory with $\order{(q^2)}$ soft behavior.
From the previous discussion, we find that the DBI action is the unique theory that has such a behavior~\cite{Cheung:2014dqa,Cheung:2016drk}. All the amplitudes of the Goldstone boson $\pi$ can be obtained from original DBI theory (with $\langle \phi\rangle =0$) by replacing $p_i\cdot p_j$ with $\tilde{p}_i\cdot \tilde{p}_j$.

Note that this feature is far from obvious if one expands the original action in \Eq{eq:DBIAction} around the boost-breaking vacuum in \Eq{eq:DBIsymmetry}. A nontrivial field redefinition is needed to render the action of $\pi$ into the form of DBI action~\cite{Grall:2020ibl}. But using on-shell construction, we see that it is a natural solution to the soft theorem. In addition, it also proves the uniqueness of DBI Inflation under such an emergent Lorentz invariance.

For the above reason, DBI Inflation corresponds to a special point in the space of inflationary models and therefore also the space of non-Gaussian statistics. This is well-known for the three and four point functions of the primordial density fluctuations. Our observation about the soft-limit of the scattering amplitude in DBI could be useful in bootstrapping the predictions of DBI for higher-point cosmological correlators.

\subsection{Coupling to Gravity}

One goal of studying amplitudes in the EFT of Inflation would be to understand inflation itself.  One might conclude that the scattering amplitudes are a valid probe of inflation in the sub-horizon regime where the geometry is approximately flat\footnote{Although, one may worry that the soft limit, $q\to 0$, is in tension with the sub-horizon limit.}. Yet, in the process we took the decoupling limit, $\Mpl \to \infty$ which, of course, does not hold in our universe.

It was argued in~\cite{Pajer:2020wnj} that the self-consistency of amplitudes in Lorentz violating EFTs only holds for flat space with no gravitational interactions.  In particular, they argue that coupling to gravity forbids the scalar amplitudes we discuss in the previous section and thus cannot be arrived at as a limit of the EFT of Inflation. If we wish to apply our soft theorems to inflation itself, it is therefore essential that we understand any potential limitations of this kind.  We will show that there is no contradiction introduced by coupling to gravity, or at least not one that is visible in the four point amplitude for the production of a graviton, $\varphi\varphi\to \varphi\gamma$.

We can understand the problem, and its resolution, most straightforwardly by considering a spectator scalar field, $\varphi$, with a Lorentz violating interaction $\lambda \dot \varphi^3$. If we include the relativistic graviton coupling that arises from the canonical kinetic term, then our action would take the form
\beq
{\cal L} \supset \frac{1}{2} \partial_\mu \varphi \partial^\mu \varphi + \frac{\lambda}{3!} \dot\varphi^3 + \frac{1}{2 \Mpl} \gamma^{\mu \nu}\partial_\mu \varphi \partial_\nu \varphi
\eeq
where we used $g_{\mu\nu} = \eta_{\mu\nu} +\Mpl^{-1}\gamma_{\mu \nu}$ with $\gamma_\mu^\mu= 0$ and kept only the leading gravitational term.  If we compute the $\varphi\varphi\to \varphi\gamma$ amplitude, we get
\beq
{\cal A}(p_{1,\varphi},p_{2,\varphi},p_{3,\varphi},p_{4,\gamma}) = -i\frac{\lambda}{\Mpl} \left( \frac{E_1 E_2 E_{12}}{2 p_1\cdot p_2} p_3^\mu p_3^\nu+\frac{E_1 E_3 E_{13}}{2 p_1\cdot p_3} p_2^\mu p_2^\nu+\frac{E_2 E_3 E_{23}}{2 p_2\cdot p_3} p_1^\mu p_1^\nu \right) \epsilon_{\mu \nu}(p_4)
\eeq
where $E_{ij} = E_i+E_j$ and $\epsilon_{\mu \nu}(p_4)$ is the polarization tensor for the graviton.  In writing this expression, we used $(p_i + p_4)^\mu \epsilon_{\mu \nu}(p_4) = p_i^\mu \epsilon_{\mu \nu}(p_4)$.  Normally one would prefer to use spinor-helicity variables for the graviton amplitude, but hopefully our reason for avoiding them will be clear below.  

Now suppose we perform a gauge transformation,
\beq
\epsilon_{\mu \nu}(p_4) \to \epsilon_{\mu \nu}(p_4) +i p_4^{\mu} f^\nu(p_4)+i p_4^{\nu} f^\mu(p_4)
\eeq
for some unknown set of functions $f^\mu(p_4)$. The amplitude will shift by ${\cal A}(p_{1,\varphi},p_{2,\varphi},p_{3,\varphi},p_{4,\gamma}) \to {\cal A}(p_{1,\varphi},p_{2,\varphi},p_{3,\varphi},p_{4,\gamma}) + \delta {\cal A}(p_{1,\varphi},p_{2,\varphi},p_{3,\varphi},p_{4,\gamma})$.
\beq
\delta {\cal A}(p_{1,\varphi},p_{2,\varphi},p_{3,\varphi},p_{4,\gamma}) = \frac{\lambda}{\Mpl} \left(E_1 E_2 E_{12}\, p_3^\mu+E_1 E_3 E_{13} \, p_2^{\mu} +E_2 E_3 E_{23} \, p_1^{\mu} \right) f_\mu(p_4) \ ,
\eeq
 Given that $g^{\mu \nu} \epsilon_{\mu \nu}(p_4) =0$, we can remove a term proportional to $p_4^\mu f_\mu (p_4)$ by subtracting a term proportional to $g^{\mu \nu}$ from polarization-stripped amplitude ${\cal A} (p_{1,\varphi},p_{2,\varphi},p_{3,\varphi},p_{4,\gamma})$ before the gauge transformation. We then notice that we can write $\delta {\cal A}$ as 
\beq
\delta {\cal A}(p_{1,\varphi},p_{2,\varphi},p_{3,\varphi},p_{4,\gamma})= -\frac{\lambda}{\Mpl} \left( E_1 E_2 p_3^{\mu}+ E_1 E_3 p_2^{\mu}+ E_2 E_3 p_1^{\mu} \,  \right) E_4 f_\mu(p_4) - E_1 E_2 E_3 p_4^{\mu} f_\mu(p_4)\ ,
\eeq
so that the final term can be removed without changing the amplitude.  However, the first term does not vanish and reflects a failure of the Ward identify for the graviton\cite{Weinberg:1964ew}. This failure of the Ward identity also implies that the amplitude written in spinor-helicity variables is not well defined, as was observed in~\cite{Pajer:2020wnj}.

The origin of this problem, is that our action is not diffeomorphism invariant. As gravity is gauging a local Lorentz transformation, it is therefore not surprising that the gravitational scattering is inconsistent if Lorentz invariance is explicitly broken. Following the construction of the EFT of Inflation, we can make the action invariant by introducing the Goldstone boson
\bea
{\cal L} &=& \frac{1}{2} \partial_\mu \varphi \partial^\mu \varphi + \frac{\lambda}{3!} (g^{\mu\nu}\partial_\mu(t+\pi) \partial_{\nu}\varphi)^3 +\frac{1}{2 \Mpl} \gamma^{\mu \nu}\partial_\mu \varphi \partial_\nu \varphi \\
&=& \frac{1}{2} \partial_\mu \varphi \partial^\mu \varphi + \frac{1}{3!}\lambda \dot\varphi^3 + \frac{1}{2\Mpl}\gamma^{\mu \nu}\partial_\mu \varphi \partial_\nu \varphi+ \frac{\lambda}{2 \Mpl} \gamma^{\mu 0}\partial_\mu \varphi \dot \varphi^2 + \frac{ \lambda}{2} \partial_\mu \pi \partial^\mu \varphi \dot \varphi^2 \ ,
\eea
where in the second line we dropped interactions that contribute only higher multiplicity or loop amplitudes.  We now see there are two additional contact contributions to the amplitude
\begin{align}
{\cal A}(p_{1,\varphi},p_{2,\varphi},p_{3,\varphi},p_{4,\gamma^{\mu 0}}) &= -i\frac{ \lambda}{M_{\rm pl}}(p_1^{\mu} E_2 E_3 + p_2^{\mu} E_1 E_3 + p_{3}^\mu E_1 E_2) \epsilon_{\mu 0} (p_4) \\
{\cal A}(p_{1,\varphi},p_{2,\varphi},p_{3,\varphi},p_{4,\pi}) &=\lambda(p_1^{\mu} E_2 E_3 + p_2^{\mu} E_1 E_3 + p_{3}^\mu E_1 E_2) p_{4,\mu}
\end{align}
Finally, we need to determine the relationship between $\pi$ and $\gamma^{0\mu}$, which is determined by the EFT of Inflation.  So to simplify the discussion, we will assume $c_s=1$ for $\pi$ so that $\pi$ and $\gamma$ propagate at the same speed. The kinetic term for $\pi$ introduces a metric coupling,
\beq
{\cal L} = \frac{1}{2}  (g^{\mu \nu} \partial_\mu(t+\pi) \partial_\nu(t+\pi)-1) = \frac{1}{2} \partial_\mu \pi \partial^\mu \pi + \frac{1}{\Mpl} \gamma^{0 \mu}\partial_\mu \pi + {\rm total \, derivative} \ .
\eeq
In addition, under a diffeomorphism, $\pi(p_4) \to \pi(p_4) - f^0(p_4)$.  Putting this altogether, the amplitude shifts by
\beq
\delta {\cal A}(p_{1,\varphi},p_{2,\varphi},p_{3,\varphi},p_{4,\gamma})+\delta {\cal A}(p_{1,\varphi},p_{2,\varphi},p_{3,\varphi},p_{4,\gamma^{\mu 0}})+ \frac{1}{\Mp} \delta{\cal A}(p_{1,\varphi},p_{2,\varphi},p_{3,\varphi},p_{4,\pi})
\eeq
under a diffeomorphism and dropping terms proportional to $p_4^\mu$ we have
\begin{align}
\delta {\cal A}(p_{1,\varphi},p_{2,\varphi},p_{3,\varphi},p_{4,\gamma}) &=-\frac{\lambda}{\Mpl} \left( E_1 E_2 p_3^{\mu}+ E_1 E_3 p_2^{\mu}+ E_2 E_3 p_1^{\mu} \,  \right) E_4 f_\mu(p_4) \\  \delta {\cal A}(p_{1,\varphi},p_{2,\varphi},p_{3,\varphi},p_{4,\gamma^{\mu 0}}) &= \frac{ \lambda}{M_{\rm pl}}(p_1^{\mu} E_2 E_3 + p_2^{\mu} E_1 E_3 + p_{3}^\mu E_1 E_2) (p_{4,\mu} f_0(p_4) +E_4 f_\mu(p_4) )\\
 \delta {\cal A}(p_{1,\varphi},p_{2,\varphi},p_{3,\varphi},p_{4,\pi})&= - \frac{ \lambda}{\Mpl}(p_1^{\mu} E_2 E_3 + p_2^{\mu} E_1 E_3 + p_{3}^\mu E_1 E_2) p_{4,\mu} f_0(p_4) \ .
\end{align}
With all three contributions, we see that $\delta {\cal A} = 0$ under diffeomorphisms, as required.

The general constraints on couplings of gravity to a theory with spontaneously broken Lorentz invariance are beyond the scope of this work.  However, we see that the most basic constraint is that Lorentz invariance is spontaneously broken (and then weakly gauged by gravity). At the same time, deriving additional constraints on the EFT from the consistency of graviton amplitudes is plausible but would likely require a more suitable treatment of the spinor helicity variables. 

\subsection{Cosmological Correlators}

The physics of inflation is encoded in cosmological correlators: equal time in-in correlation functions calculated around the quasi-de Sitter background that describes inflation.  The EFT of Inflation is particularly useful in characterizing non-Gaussian cosmological correlators.  Concretely, for single-field inflation, the scalar metric fluctuation, $\zeta$, eats the Goldstone such that $\zeta \approx - H \pi$\cite{Maldacena:2002vr} outside the horizon.  With the additional assumption of time-independent couplings, the bispectrum gives the leading in-in~\cite{Weinberg:2005vy,Weinberg:2006ac} non-Gaussian correlator, which is given by \cite{Chen:2006nt,Cheung:2007sv,Pajer:2020wxk}
\beq\label{eq:bispectrum}
\langle \zeta(\k_1) \zeta(\k_2) \zeta(\k_3) \rangle' =\Delta_\zeta^{4} \frac{12 \frac{g_3 c_s^2}{f_\pi^4} e_{3}^{2}- \frac{\delta_c}{2 f_\pi^4} \left(-4 k_{T} e_{2} e_{3}-4 k_{T}^{2} e_{2}^{2}+11k_{T}^{3} e_{3}-3 k_{T}^{4} e_{2}+ k_{T}^{6}\right)}{e_{3}^{3} k_{T}^{3}} \ ,
\eeq
where $\Delta_\zeta = H^2 / (\sqrt{2} f_\pi^2)$, we dropped terms that are slow-roll suppressed, and defined the symmetric polynomials of $k_i$ in terms of the total energy $k_T = k_1+k_2+k_3$, $e_2 = k_1 k_2 +k_1 k_3+k_2 k_3$ and $e_3= (k_1 k_2 k_3)$.

One of the key concepts of the cosmological bootstrap is the idea that the residue of the total energy pole (i.e.~the leading pole when we analytically continue to $k_T \to 0$) is the flat-space scattering amplitude~\cite{Maldacena:2011nz,Raju:2012zr}. Indeed, in this case we see that the leading behavior, $\Delta_\zeta^4 36 g_3 c_s^2 (k_1 k_2 k_3)/(k_1 k_2 k_3 k_T^3)$ contains the amplitude ${\cal A}_3 = -6 i g_3 E_1 E_2 E_3 $ after identifying $k_i \to E_i$ for a massless particle.  However, we can tune $g_3\to 0$ such that the on-shell three point amplitude vanishes, ${\cal A}_3 =0$. In fact, in flat space the field redefinition $\pi \to \pi - \delta_c \dot \pi \pi$ removes the $\delta_c \dot \pi \partial_\mu \pi \partial^\mu \pi$ interaction from the action.  Yet, the bispectrum contains a number of terms that appear to follow from this interaction, including several with poles at $k_T \to 0$.

The resolution of this tension is that the field redefinition doesn't remove all the interactions in the inflationary background~\cite{Grall:2020ibl}.  The action in an FLRW background, $S = \int d^4 x a^3(t) {\cal L}$ with the same Lagrangian
\beq
{\cal L} = \frac{1}{2} \partial_\mu \pi \partial^\mu \pi + \frac{\delta_c}{2} \dot \pi \partial_\mu \pi \partial^\mu \pi + g_3 \dot \pi^3 \ .
\eeq
Performing the field redefinition $\pi \to \pi - \delta_c \dot \pi \pi$ gives the shift of the action
\beq
{\cal L} \to {\cal L} - \delta_c \dot \pi \partial_\mu \pi \partial^\mu \pi - \delta_c \pi \partial_\mu \dot \pi \partial^\mu \pi + {\cal O}(\pi^4) \ .
\eeq
In flat space, these two terms are related by a total derivative.  However, in our FLRW background the total derivative takes the form
\beq
\frac{d}{dt} (a^3 \pi \partial_\mu \pi \partial^\mu \pi ) =a^3 \left( \dot \pi \partial_\mu  \pi \partial^\mu \pi  + 2  \partial_\mu \dot \pi \partial^\mu \pi  + 3 H c_s^{-2}\pi \dot \pi^2 -a H \pi \partial_i \pi \partial^i \pi\right)
\eeq
As a result, after the field redefinition we have 
\beq
{\cal L}\to \frac{1}{2} \partial_\mu \pi \partial^\mu \pi +  \frac{\delta_c}{2}\left(3 H c_s^{-2}\pi \dot \pi^2 - H \pi a^{-2} \partial_i \pi \partial^i \pi\right) + g_3 \dot \pi^3 \ .
\eeq
Given that $\dot \pi\pi \to 0$ as $a(t) \to \infty$, we can still calculate the contribution to the scalar metric fluctuation with $\zeta = -H \pi$.  Using these three operators we have three contributions to the bispectrum \cite{Cheung:2007sv}
\begin{align}
B_{\dot \pi^3} &= \Delta_\zeta^4 \frac{ g_3 \, c_s^2}{f_\pi^4} \frac{12 e_3^2}{e_3^3 k_T^3} \\
B_{\pi\dot \pi^2} &=  \Delta_\zeta^4 \frac{\delta_c}{f_\pi^4} \frac{-6 e_3 k_T^3 + 3 k_T^2 e_2^2+ 3 k_T e_2 e_3 }{e_3^3 k_T^3}\\
B_{\pi \partial_i\pi \partial^i \pi} &= \Delta_\zeta^4 \frac{ \delta_c}{2 f_\pi^4} \frac{ e_3 k_T^3-2 k_T^2 e_2^2-2 k_T e_3 e_2 +3 k_T^4 e_2  - k_T^6}{e_3^3 k_T^3}
\end{align}
Combining these terms we reproduce the bispectrum in Equation~(\ref{eq:bispectrum}).

Interestingly, when we set $g_3 = 0$, the flat-space amplitude vanishes but there remains a total-energy pole in the correlator~\cite{Grall:2020ibl}, which arises in DBI Inflation for example. However, this is a result of how we take the flat-space limit.  Concretely, the interaction $\pi \dot \pi^2$ also contributions a non-zero amplitude in flat-space. This interaction is present after our field redefinition in dS, but would vanish in the $H \to 0$ limit. However, this does not imply a suppression of the cosmological correlator as the powers of $H$ are fixed by dimensional analysis.  The end result is that there remains a total energy pole when $g_3 =0$ associated with the $\pi \dot \pi^2$ amplitude.

\subsection{Multifield and Quasi-Single Field Inflation}

The unique relationship between soft theorems and the predictions of inflation are specific to single-field inflation.  Concretely, when the dynamics of inflation are controlled by a single degree of freedom, we can always choose a gauge where that degree of freedom is the scalar mode of the metric, or equivalently, the goldstone boson $\pi$. However, in the presence of multiple degrees of freedom, the relationship between the observable scalar fluctuation and the goldstone boson $\pi$ is not longer fixed by diffeomorphism invariance.  

In the presence of multiple light scalar fields, we can pick a gauge where fluctuations along the inflationary trajectory are given by the goldstone, $\pi$ and all additional fields, $\vec \phi$ are transverse. During inflation, $\pi$ is eaten by the metric and the $\vec \phi$ fields are effectively isocurvature modes (at least at linear order).  However, the dynamics of the inflationary or post-inflationary universe can convert the isocurvature fluctuations into metric fluctuations.  Such processes are local on the scales we will observe and therefore $\zeta = F[\pi, \vec \phi]$ for some model-dependent function $F$~\cite{Senatore:2010wk}. One can easily arrange models where $\zeta$ is determined by a single transverse scalar, $\zeta \approx \kappa \phi$ so that the statistics of the metric fluctuations are not fixed by $\pi$ or the Ward identities discussed here.

Quasi-single field inflation~\cite{Chen:2009zp,Baumann:2011nk,Chen:2012ge,Noumi:2012vr}, also known as cosmological collider physics~\cite{Arkani-Hamed:2015bza,Meerburg:2016zdz,Alexander:2019vtb,Kumar:2019ebj,Wang:2019gbi,Bodas:2020yho}, provides an interesting middle ground where additional fields are important but do not destroy the relationship between the metric fluctuations and the Goldstone boson of the EFT of Inflation.  Additional massive fields, $m ={\cal O }(H)$, and particles with spin will decay outside the horizon.  They do not survive until the end of inflation and therefore reheating and subsequent evolution is determined solely by $\pi$ via $\zeta \approx - H \pi$.  These particles can still alter the statistics of $\pi$, and therefore $\zeta$, through interactions during inflation. These interactions are governed by the EFT of Inflation coupled to additional matter and are subject to the same mixed constraints discussed in Section~\ref{sec:mixed}.

\section{Outlook and Conclusions}
\label{sec:conclusions}

The structure of cosmological correlators is deeply tied to scattering amplitudes in flat space. At the level of a Lagrangian, this may seem like a vacuous statement, as both the amplitudes and correlators can be determined via the Feynman rules.  Yet, amplitudes are known to display a wide range of constraints and simplifications that are hardly apparent from the Lagrangian. This same simplicity is hiding in cosmological correlators, which are known to contain the amplitude as the residue of the total energy pole.  In recent years, this relationship has inspired the cosmological bootstrap program~\cite{Baumann:2022jpr}, which aims to understand the structure and consistency of our cosmological observables without directly appealing to the Lagrangian or the Feynman rules.

A unique challenge of this program is that inflation is described by a non-relativistic EFT~\cite{Creminelli:2006xe,Cheung:2007st}. If amplitudes serve as a model for our understanding of cosmology, then it is noteworthy that we understand considerably less about amplitudes when time translation and/or boosts are broken. Our goal in this paper was to understand how the structure of the EFT of Inflation is reflected in the scattering amplitudes.  As inflation spontaneously breaks Lorentz boosts, the constraints from Lorentz invariance manifest themselves in Ward identities that connect the soft Goldstone boson emission to the hard scattering process.

The self-consistency of amplitudes is known to place non-trivial constraints on the space of EFTs and their Wilson coefficients~\cite{deRham:2022hpx}. Famously some higher-derivative couplings are forced to be positive~\cite{Pham:1985cr,Adams:2006sv}.  In addition, consistency of the soft theorems may determine the full structure of the amplitude. Any hope of deriving similar results in the inflationary context relies on a deeper understanding of amplitudes with broken Lorentz boosts~\cite{Baumann:2015nta}.  As we have seen, this is particularly challenging for couplings to the graviton, as it gauges the broken symmetries. Yet, even for scalars, our limited understanding of the analytic structure of the amplitude and the absence of crossing symmetry are clear obstacles to using flat-space techniques directly (see e.g.~\cite{Baumann:2019ghk,Grall:2021xxm,Creminelli:2022onn} for recent progress).  

There are several avenues for future work. A natural next step for this work would be to apply the soft theorem to derive recursion relations for on-shell amplitudes~\cite{Cheung:2015ota,Luo:2015tat,Bartsch:2022pyi}.
Since the inflaton we consider is derivatively coupled, we need to have full control of the $\order(q)$ soft behavior in order to derive the recursion relations. The soft theorem we derived here only applies the spatial component of the soft momentum $q$.
The soft behavior of the energy component depends both on lower order scattering amplitudes and a new coupling constant $M_n^4$ for an $n$-particle amplitude. One needs to disentangle the two contributions in order to derive the recursion relations.

As the Goldstone boson naturally mixes with graviton, it would be interesting to extend the soft theorem to include gravitons. We expect that gravitons and Goldstone bosons are closely related since the gauge invariance is only restored after considering the combination of the two. 
Their dependence could lead to a ``transmutation'' at the level of on-shell amplitudes~\cite{Cheung:2017ems}.
It would also be great to understand the soft theorems of derivatively-coupled Goldstone bosons in terms of geometric perspective~\cite{Cheung:2021yog,Cheung:2022vnd,Cohen:2022uuw}. 

For cosmological applications, one would like to explore the relationship between soft theorems in cosmology and soft theorems in amplitudes~\cite{Mirbabayi:2016xvc}, particularly in the context of multi-field inflation. It would also be interesting to investigate the full soft theorem for loop amplitudes; extending the non-perturbative statements to the $c_s \le 1$ case would be important for a large class of inflation models. These results could sharpen our understanding of amplitudes in the context of inflation, with the hope to place stringent constraints on the candidates of inflationary models~\cite{deRham:2022hpx}.

Finally, while our work focuses on the EFT of Inflation, there is a zoo of EFTs for spontaneous breaking of Lorentz boosts, such as framid, phonons, and Galileid~\cite{Nicolis:2015sra}. For theories with enhanced Adler zeros, the on-shell methods naturally unify different EFTs and impose sharp constraints~\cite{Cheung:2016drk}. It would be fascinating to unify all boost-breaking EFTs also from the on-shell perspective.

\paragraph{Acknowledgements} \hskip 5pt We are grateful to Daniel Baumann, Tim Cohen,
Clifford Cheung, Nathaniel Craig, Maria Derda, Andreas Helset, Austin Joyce, Aneesh Manohar, Julio Parra-Martinez, Riccardo Penco, Akhil Premkumar, and the participants of the Simons Symposium {\it Amplitudes meets Cosmology} for helpful discussions.
CHS thanks Aneesh Manohar for teaching him nuclear physics.
DG and YH are supported by the US~Department of Energy under Grant~\mbox{DE-SC0009919}.
CHS is supported in part by the U.S.\ Department of Energy (DOE) under award number~DE-SC0009919 and the National Science Foundation (NSF) under Grant No. NSF PHY-1748958. This work was completed at the Aspen Center for Physics, which is supported by the NSF grant PHY-1607611.

\clearpage
\phantomsection
\addcontentsline{toc}{section}{References}
\bibliographystyle{utphys}
\bibliography{Wardrefs}

\end{document}